\documentclass[useAMS,usenatbib]{mn2e} 
\usepackage{graphics, graphicx} 

\title[Polarisation profiles of southern pulsars at 3.1
GHz]{Polarisation profiles of southern pulsars at 3.1 GHz}

\author[Karastergiou et al.]  {A.~Karastergiou$^1$, S.~Johnston$^1$ \& R.~N.~Manchester$^2$\\
$^1$School of Physics, University of Sydney, NSW 2006, Australia\\
$^2$Australia Telescope National Facility, CSIRO, P.O. Box 76, Epping, NSW 1710, Australia\\
}

\date{Released 2004 Xxxxx XX}
 
\pagerange{\pageref{firstpage}--\pageref{lastpage}} \pubyear{2004} 
 
\def\LaTeX{L\kern-.36em\raise.3ex\hbox{a}\kern-.15em 
    T\kern-.1667em\lower.7ex\hbox{E}\kern-.125emX} 
 
\begin{document} 
 
\label{firstpage} 
 
\maketitle 
 
\begin{abstract} 
We present polarisation profiles for 48 southern pulsars observed with
the new 10-cm receiver at the Parkes telescope. We have exploited the
low system temperature and high bandwidth of the receiver to obtain
profiles which have good signal to noise for most of our sample at
this relatively high frequency.  Although, as expected, a number of
profiles are less linearly polarised at 3.1 GHz than at lower
frequencies, we identify some pulsars and particular components of
profiles in other pulsars which have increased linear polarisation at
this frequency.  We discuss the dependence of linear polarisation with
frequency in the context of a model in which emission consists of the
superposition of two, orthogonally polarised modes. We show that a
simple model, in which the orthogonal modes have different spectral
indices, can explain many of the observed properties of the frequency
evolution of both the linear polarisation and the total power, such as
the high degree of linear polarisation seen at all frequencies in some
high spin-down, young pulsars. Nearly all the position angle profiles
show deviations from the rotating vector model; this appears to be a
general feature of high-frequency polarisation observations.
\end{abstract} 
 
\begin{keywords} 
pulsars: general - polarisation
\end{keywords} 
%
\section{Introduction}
The physics of pulsar magnetospheres can be tackled via two
observational routes. Observing single pulses from radio pulsars gives
information on the instantaneous plasma conditions, the production of
linear and circular polarisation and radiation mechanism. The
resulting phenomenology and the underlying physics have been discussed
recently for giant pulses \citep{jr03}, drifting subpulses
\citep[e.g.][]{es02,es03b,cr04} and circular polarisation effects
\citep{kj04,ml04a}. However, current sensitivity permits such
observations only on a small sample of the total pulsar population,
especially at frequencies above about 1~GHz.

\begin{table}
\begin{tabular}{ccccccc}
Pulsar & Period & $S_{1400}$ & $S_{3100}$ & $\langle L\rangle$/S &
$W_{50}$ & $W_{10}$\\
 & (msec) & (mJy) & (mJy) & (\%) & ($\degr$) & ($\degr$) \\
\hline                                          
J1012$-$5857  &   819  &     1.7  & 0.6 & 13 & 3 & 9  \\
J1038$-$5831  &   661  &     0.8  & 0.6 & 26 & 3 & 12 \\
J1048$-$5832  &   123  &     6.5  & 2.9 & 80 & 13 & 29 \\
J1110$-$5637  &   558  &     1.8  & 0.7 & 15 & 13 & 16 \\
J1114$-$6100  &   880  &     2.0  & 2.1 & -  & 8  & 15 \\
J1126$-$6054  &   202  &     1.0  & 0.3	& 20 & 7 & 17 \\
J1133$-$6250  &   1022 &     2.9  & 1.5 & - & 91 & 112 \\
J1302$-$6350  &   48   &     3.6  & 2.8 & 97 & 256 & 274 \\
J1306$-$6617  &   473  &     2.5  & 1.3 & 15 & 30 & 47 \\
J1319$-$6056  &   284  &     1.2  & 0.4 & 39 & 12 & 21 \\
J1327$-$6301  &   196  &     3.2  & 0.9 & 24 & 15 & 71 \\
J1338$-$6204  &   1238 &     3.8  & 2.0	& 25 & 9 & 35 \\
J1352$-$6803  &   628  &     1.1  & 0.4	& - & 18 & - \\
J1410$-$7404  &   278  & -        & 2.0 & - & 3 & - \\
J1413$-$6307  &   394  &     0.9 & 0.5	& 34 & 2 & 8\\
J1512$-$5759  &   128  &     6.0 & 1.1	& - & 12 & 26\\
J1517$-$4356  &   650  & -        & 0.3	& - & 6 & - \\
J1522$-$5829  &   395  &     4.3 & 1.4  & 33 & 13 & 25 \\
J1534$-$5405  &   289  &     1.2 & 0.4	& - & 14 & - \\
J1535$-$4114  &   432  & -        & 1.5	& 34 & 10 & 20 \\
J1539$-$5626  &   243  &     4.6 & 1.8  & 46 & 15 & 24 \\
J1611$-$5209  &   182  &     1.2 & 0.9	& - & 5 & 200 \\
J1614$-$5048  &   231  &     2.4 & 0.7  & 92 & 4 & 16 \\
J1615$-$5537  &   791  &     0.4 & 0.3  & - & 8 & - \\
J1630$-$4733  &   575  &     4.0 & 2.9  & 20 & 15 & 33 \\
J1633$-$5015  &   352  &     5.7 & 1.2  & - & 5 & - \\
J1633$-$4453  &   436  &     1.9 & 0.5  & 19 & 10 & 18 \\
J1637$-$4553  &   118  &     1.1 & 0.4  & 76 & 11 & 21 \\
J1640$-$4715  &   517  &     1.2 & 0.6	& - & 6 & 19\\
J1646$-$4346  &   231  &     1.0 & 0.4  & 50 & 11 & 21 \\
J1653$-$3838  &   305  &     1.3 & 1.2	& 37 & 4 & 22 \\
J1655$-$3048  &   542  & -       & 0.5	& -  & 55 & 72 \\
J1701$-$4533  &   322  &     2.5 & 0.6	& 25 & 18 & 24 \\
J1707$-$4053  &   581  &     7.2 & 1.7	& 35 & 11 & 18\\
J1709$-$4429  &   102  &     7.3 & 5.9  & 93 & 28 & 47 \\
J1712$-$2715  &   255  & -       & 0.9  & - & 51 & 70 \\
J1719$-$4006  &   189  &     1.1 & 0.6	& - & 14 & 22\\
J1721$-$3532  &   280  &    11.0 & 6.4	& 35 & 14 & 29\\
J1722$-$3632  &   399  &     1.6 & 0.8  & -  & 4 & 29 \\
J1733$-$3716  &   337  &     3.4 & 1.5	& 31 & 8 & 52\\
J1737$-$3555  &   397  &     0.7 & 0.5  & -  & 8 & - \\
J1742$-$4616  &   412  & -       & 0.5	& - & 29 & - \\
J1749$-$3002  &   609  &     3.7 & 1.1  & 26 & 40 & 47\\
J1750$-$3157  &   910  &     1.2 & 0.5	& - & 28 & 39\\
J1808$-$3249  &   364  & -       & 0.9  & 32 & 15 & 26 \\
J1820$-$1818  &   309  &     1.1 & 0.6  & - & 14 & -	\\
J1943+0609  &   446  & -         & 0.4  & - & 10 & - \\
J2007+0809  &   325  & -         & 2.5  & - & 68 & -	\\
\end{tabular}
\caption{The name, period, flux density, fractional linear
  polarisation and profile width of the observed pulsars. $S_{1400}$
  values (apart from J1302$-$6350) were taken from Hobbs et
  al. (2004).}
\end{table}

The second observational route comes from a study of the time-averaged
polarisation profiles in pulsars. These integrated profiles are
extremely stable and thus yield information on the long-term structure
of the magnetic field, the average properties of the magnetosphere and
the geometry of the star. For example, according to the model of
\citet{rc69a}, the position angle (PA) of the linear polarisation is
tied to the magnetic field in the vicinity of a magnetic pole and
therefore changes in a well-defined manner as the beam of the pulsar
sweeps past our line of sight. In principle, this allows the geometry
of the star to be determined, but this is difficult in practice for
two main reasons \citep{ew01}. Firstly, discontinuous jumps in PA are
observed in many pulsars; these jumps are generally close to $90\degr$
and have been associated with emission from orthogonally polarised
modes \citep{mth75,brc76,ch77,crb78}, hereafter OPM.  Single-pulse
studies, supported by the statistical model of \citet{scr+84} and
later \citet{mck97} and \citet{ms98}, showed that the observed
polarisation can be explained to some degree by superposed rays in
orthogonally polarised states \citep{kkj+02}. The superposition of
such rays results in the total intensity being the sum of the OPM
intensities and the linear polarisation being the difference in linear
polarisation in each mode. Secondly, the longitude over which emission
occurs is usually very small and additional, non-orthogonal deviations
in the PA swing are often seen. These are likely due to propagation
effects in the pulsar magnetosphere.

Average polarisation profiles have also been used to characterise the
various components which make up the pulse
profile. \citet{ran83,ran83a,ran86} used average profiles to classify
components as core or cone, which have different total power and
polarisation properties. \citet{lm88} confirmed the differences
between component types, but saw no reason to advocate different
emission mechanisms for each type. Rankin also concluded that, on
average, significant circular polarisation is only seen in core
components, a conclusion disputed by \citet{hmxq98}.  Also, single
pulse studies have since shown significant circular polarisation in
cone components \citep{kjm+03,kj04}.

The frequency dependence of the integrated profile properties has been
a research area for many years. In the southern hemisphere, the
average polarisation profiles of strong pulsars have been obtained at
400, 600 and 1612 MHz, and presented in a series of papers
\citep{hmak77,mhma78,mhm80} with later observations at 800 and 950 MHz
\citep{vdhm97}. Observations at higher frequencies have generally been
confined to pulsars visible from the northern hemisphere
\citep[e.g.][]{mgs+81,hx97,hkk98}. There are three broad
generalisations that can be made about the frequency evolution of
pulsar profiles. First, the total power profiles tend to be more
complex at high frequencies as outrider components become more
prominent compared to the central component \citep{ran83}. Secondly,
the percentage polarisation generally decreases as the frequency
increases \citep{mhm80,mgs+81}. Finally, the OPM phenomenon also
becomes more prevalent at higher frequencies \citep[e.g.][]{kkj+02},
which is thought to be the result of strong diffraction effects that
occur lower in the pulsar magnetosphere \citep[e.g.][]{pet01}.

Many new pulsars have been discovered since the end of the 1980s,
rather few of which have observed polarisation profiles.  We have
conducted observations in full polarisation of 48 southern pulsars at
3.1 GHz originating from the \cite{jlm+92} and \cite{ebvb01} surveys,
to describe total power and polarisation behaviour at this relatively
high radio frequency.  Some of the pulsars from the Johnston et
al. survey have been published in full polarisation, mostly by
\citet[hereafter QMLG]{qmlg95} at 1.4 GHz. No published polarisation
profiles of the pulsars discovered by Edwards et al. exist to date.
We draw a comparison between our current observations and the QMLG
observations where appropriate.
%
\section{Observations}

All observations were carried out with the 64-m radio telescope
located in Parkes, Australia.  A total of 50~h of observing were
obtained in the period 2004 Jan 24 to Jan 27 at a central observing
frequency of 3094~MHz. The observations were made using a dual
10/50~cm receiver (see Granet et al. 2004 for a description of the
feed).  The system has orthogonal linear feeds and cryogenically
cooled preamplifiers, giving a system equivalent flux density on cold
sky of 49~Jy.  A signal can be injected at an angle of 45$^{\circ}$ to
the feed probes for calibration purposes.

Data were recorded using the wide-band pulsar correlator. A total
bandwidth of 1024~MHz was used. Channel bandwidths were 1~MHz and
there were typically 1024 phase bins across the pulsar period. The
data were folded on-line for an interval of 60~s and written to
disk. Total integration times were either 30 or 60~min depending on
the flux density of the pulsar. These observations were made during a
phase when the correlator was still being commissioned. Unfortunately,
some of the observations suffer from artifacts caused by the
hardware. In particular, Stokes $I$ sometimes has `saw-tooth' (non
Gaussian) baselines and also shows extended `shoulders' before and
after the profile peaks as can be seen in some of the Figures
(e.g. PSR J1319$-$6056). These artifacts do not affect the other
Stokes parameters and have little overall impact on our results.

Flux calibration was obtained by observations of Hydra~A which is
assumed to have a flux density of 20.95~Jy at 3100~MHz.  Polarisation
calibration was carried out by observing a pulsed signal of known
polarisation prior to each observation of a pulsar.  Differential gain
and delay between the two feed probes could then be accounted for.
Off-line processing used the PSRCHIVE software application
\cite{hvm04} specifically written for analysis of pulsar data. Faraday
rotation measures, where known, were applied to the data after
calibration.

%
\begin{figure*}
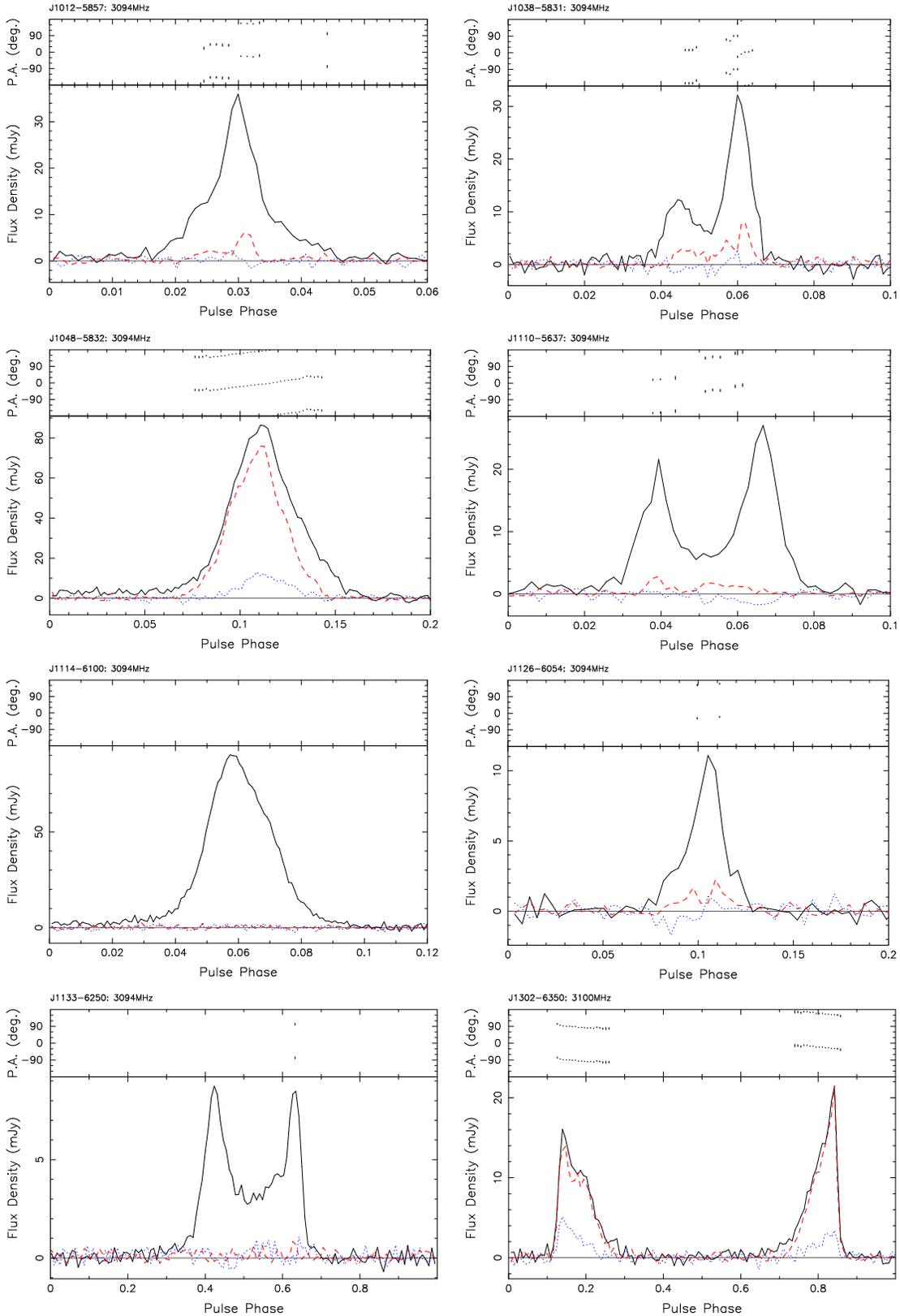

\begin{center}
\begin{tabular}{cc}
\resizebox{0.4\hsize}{!}{\includegraphics[angle=-90]{J1012-5857.ps}}&
\resizebox{0.4\hsize}{!}{\includegraphics[angle=-90]{J1038-5831.ps}}\\
\resizebox{0.4\hsize}{!}{\includegraphics[angle=-90]{J1048-5832.ps}}&
\resizebox{0.4\hsize}{!}{\includegraphics[angle=-90]{J1110-5637.ps}}\\
\resizebox{0.4\hsize}{!}{\includegraphics[angle=-90]{J1114-6100.ps}}&
\resizebox{0.4\hsize}{!}{\includegraphics[angle=-90]{J1126-6054.ps}}\\
\resizebox{0.4\hsize}{!}{\includegraphics[angle=-90]{J1133-6250.ps}}&
\resizebox{0.4\hsize}{!}{\includegraphics[angle=-90]{J1302-6350.ps}}\\
\end{tabular}
\end{center}
\caption{Polarisation profiles at 3.1 GHz. The solid line is the mean
total power, the dashed line is the average linearly polarised power
and the dotted line is the mean circular polarisation
($I_{LH}-I_{RH}$). On the top part of each plot is the position angle
(PA) profile, with points drawn wherever the linear polarisation
exceeds 5 times the noise. The PA has an arbitrary zero offset.}
\end{figure*}
\addtocounter{figure}{-1}
\begin{figure*}
\begin{center}
\begin{tabular}{cc}
\resizebox{0.4\hsize}{!}{\includegraphics[angle=-90]{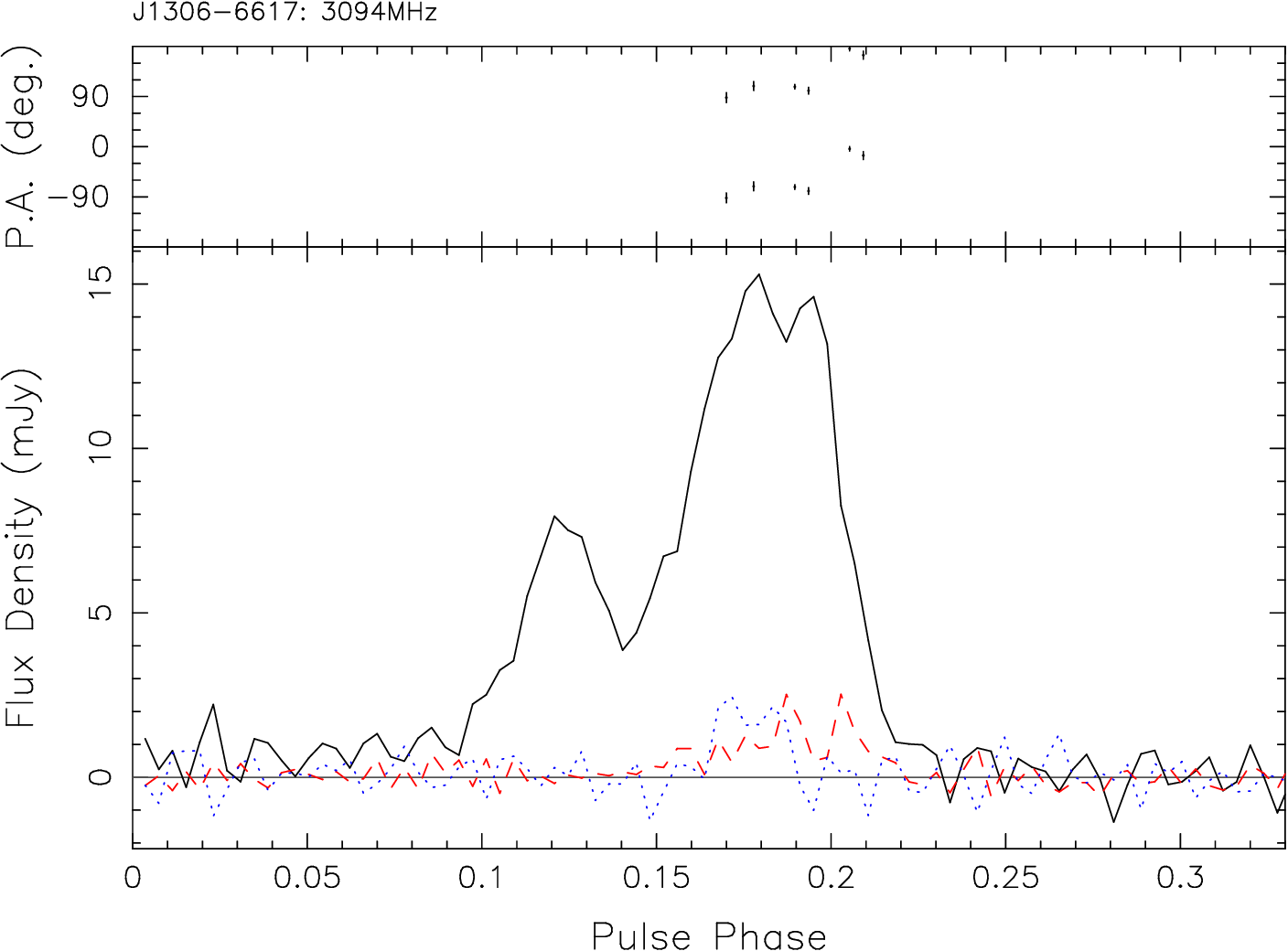}}& 
\resizebox{0.4\hsize}{!}{\includegraphics[angle=-90]{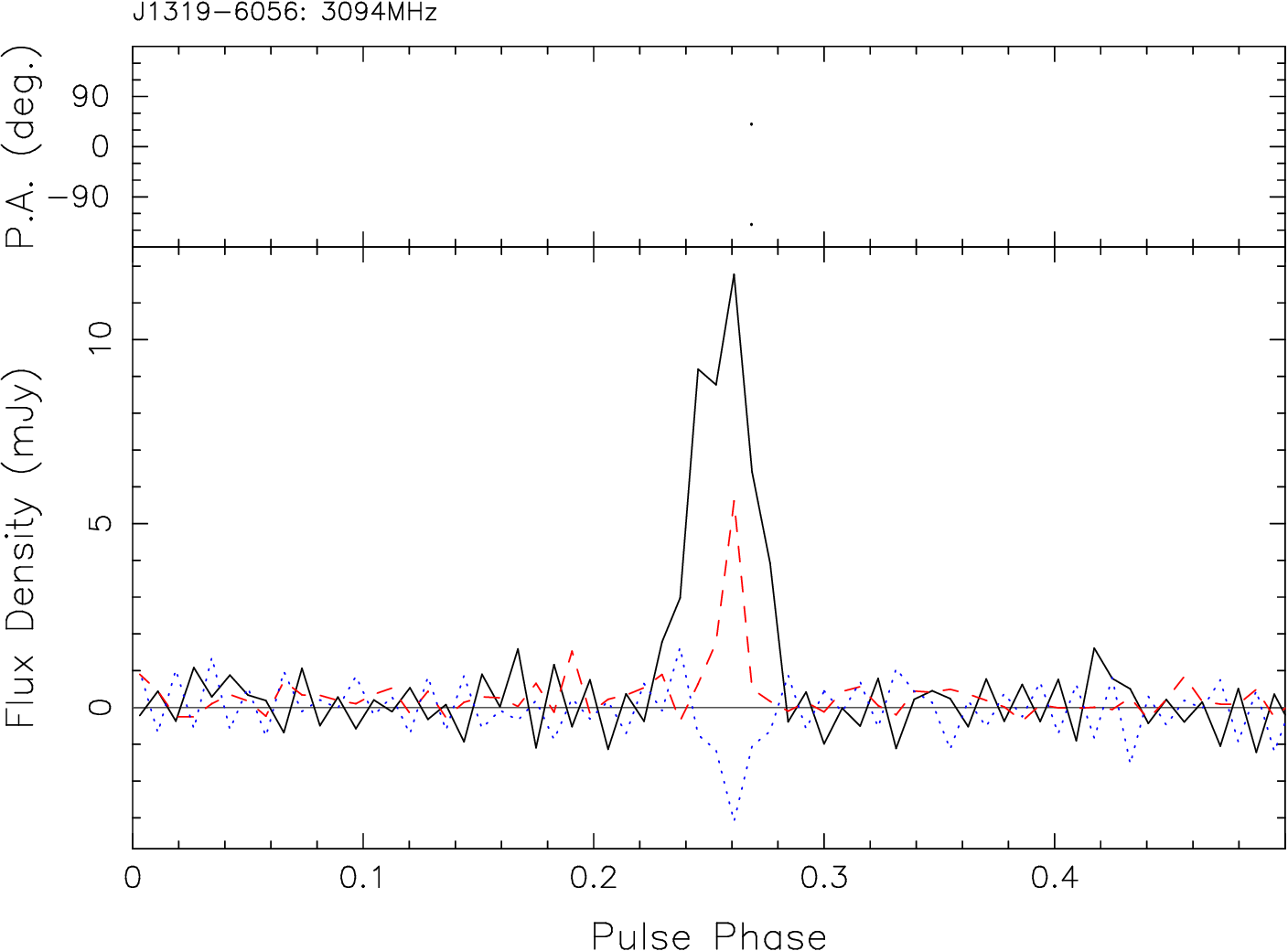}}\\
\resizebox{0.4\hsize}{!}{\includegraphics[angle=-90]{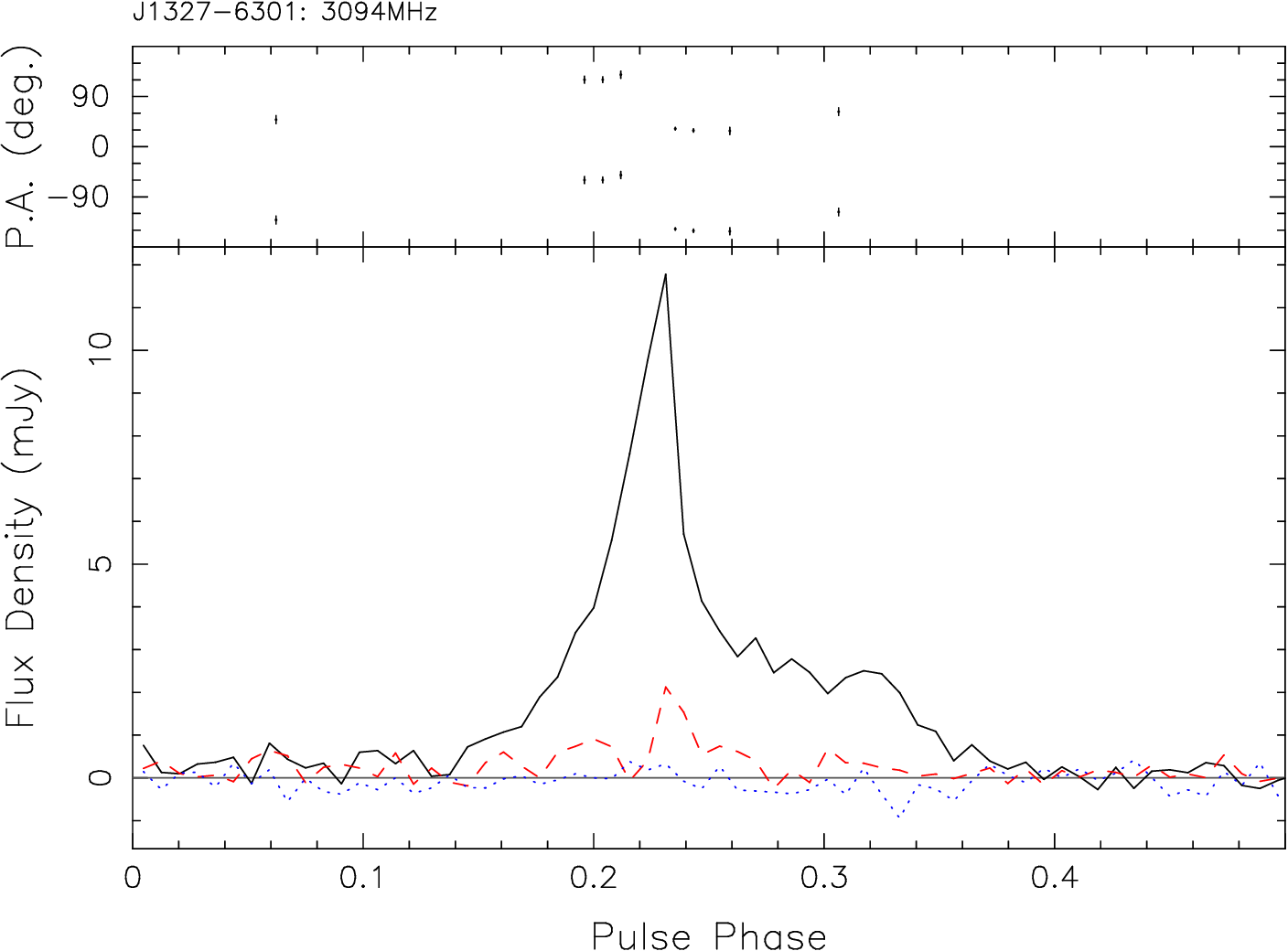}}& 
\resizebox{0.4\hsize}{!}{\includegraphics[angle=-90]{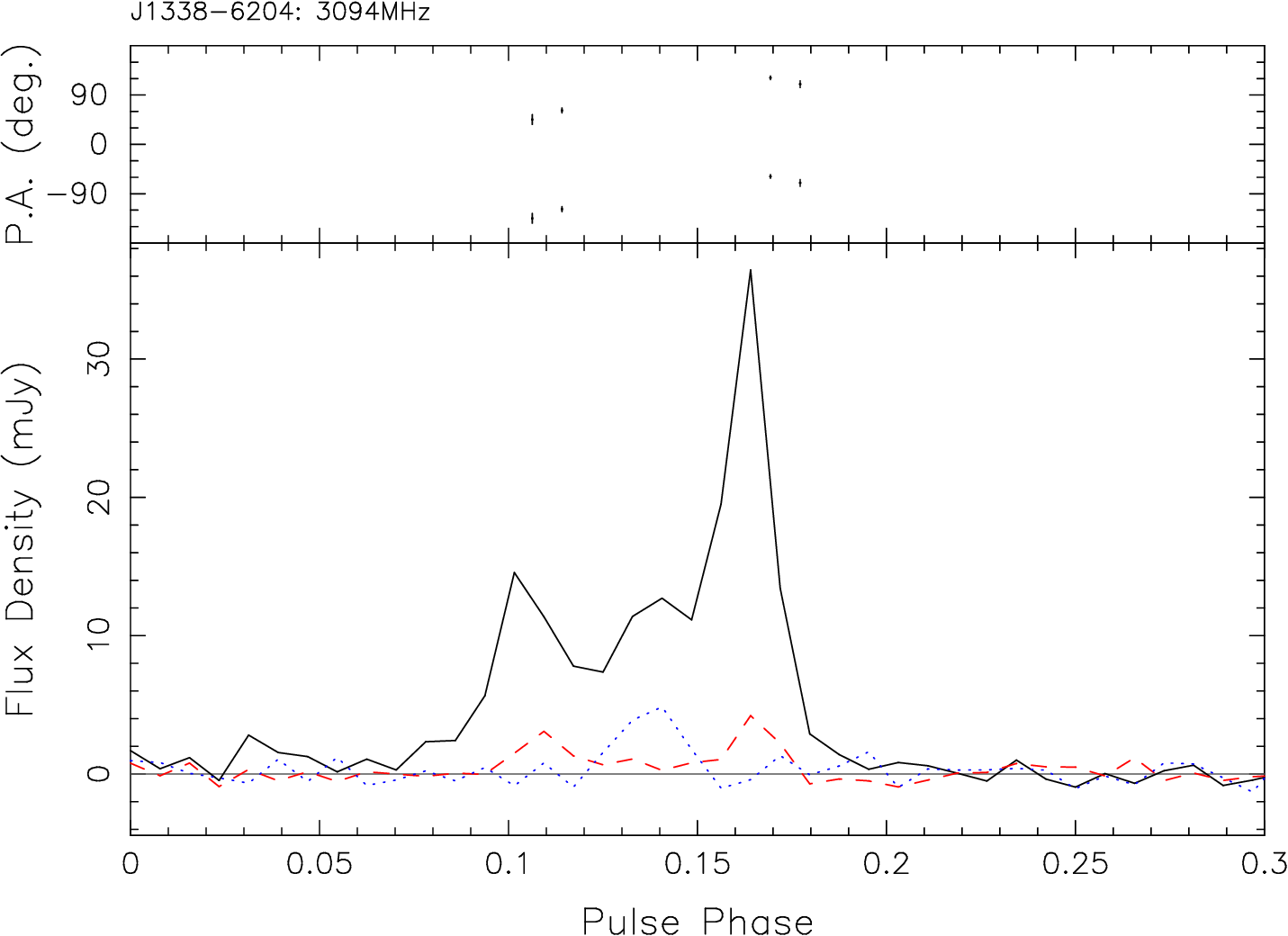}}\\
\resizebox{0.4\hsize}{!}{\includegraphics[angle=-90]{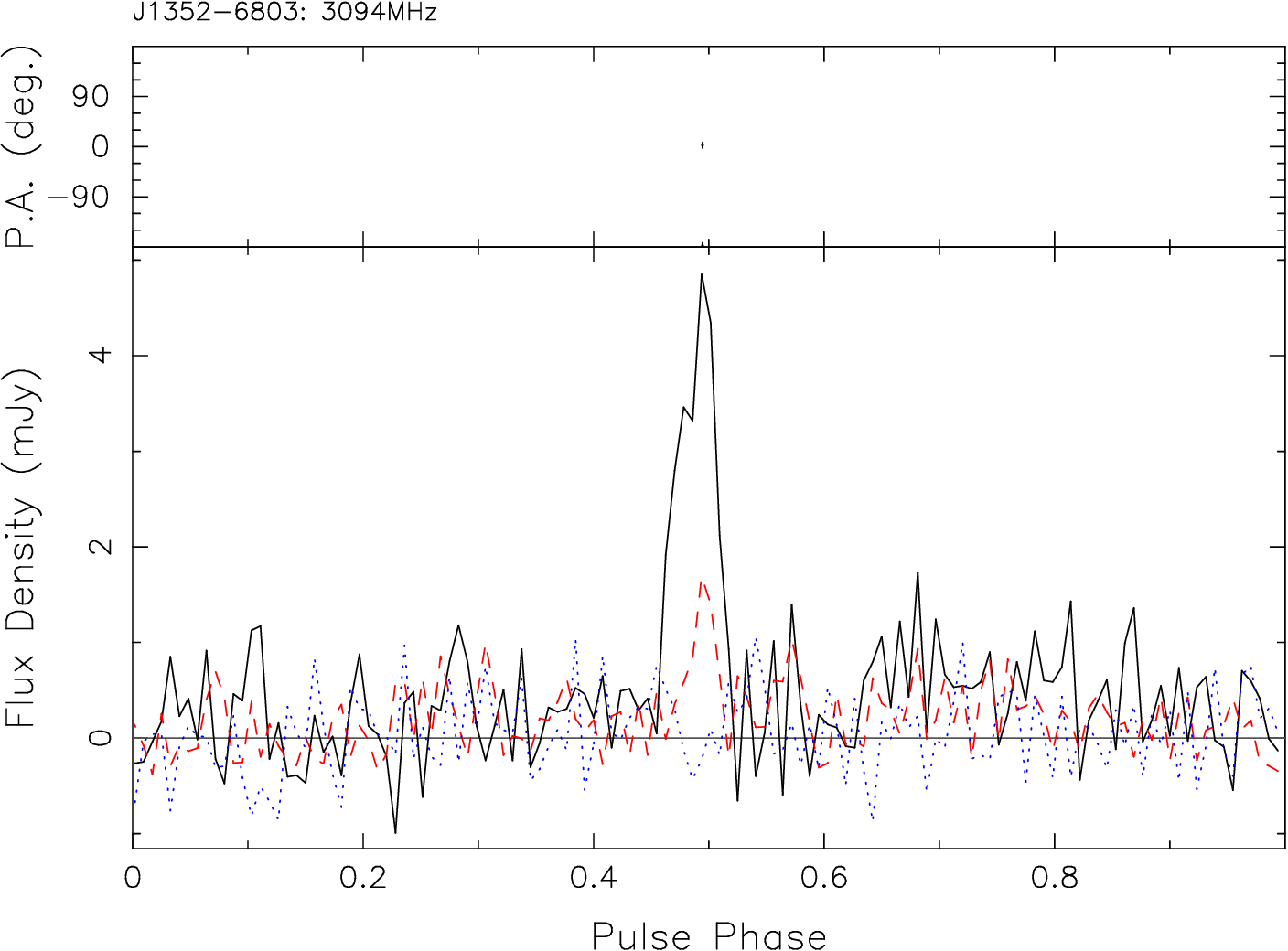}}& 
\resizebox{0.4\hsize}{!}{\includegraphics[angle=-90]{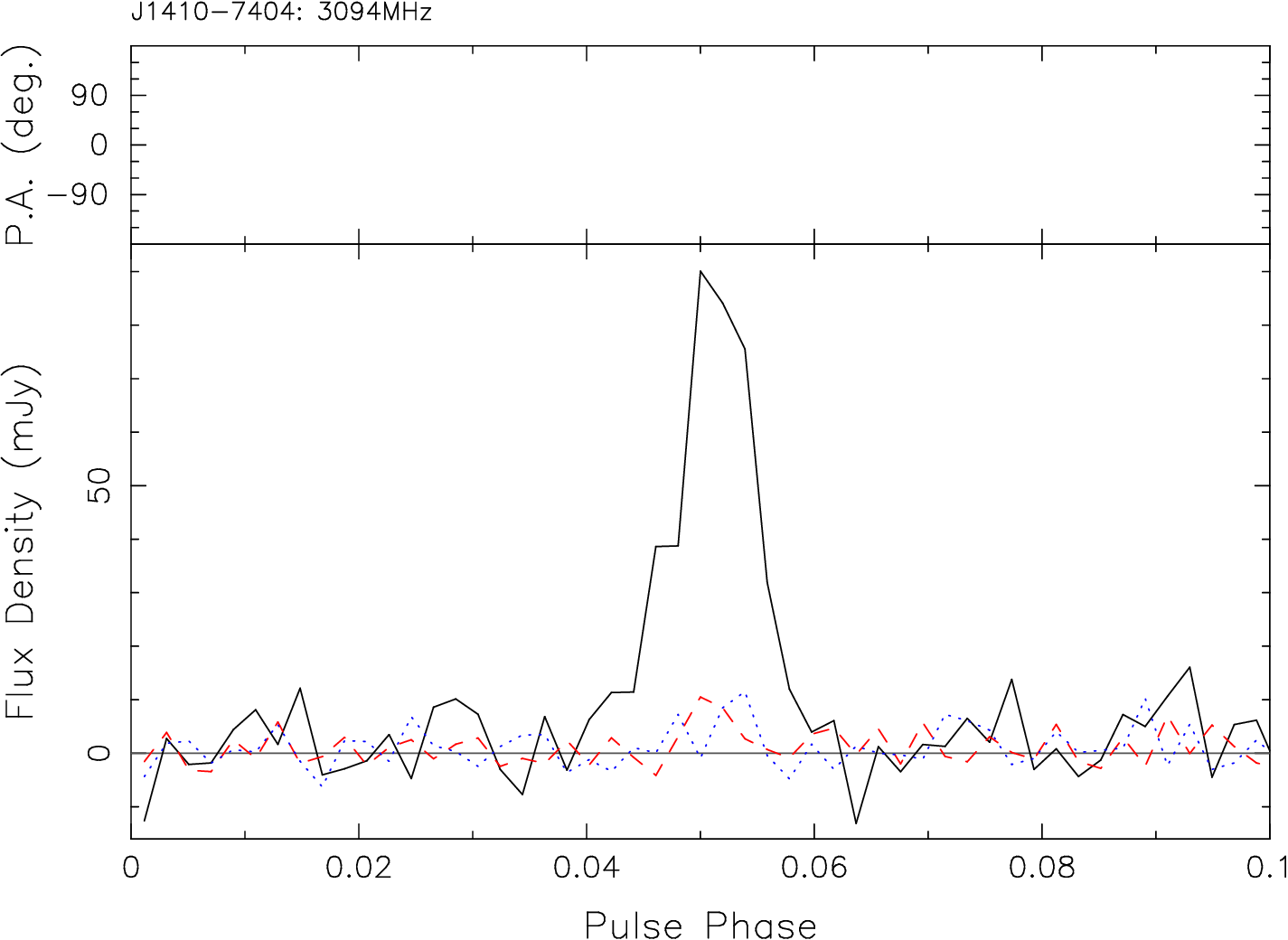}}\\
\resizebox{0.4\hsize}{!}{\includegraphics[angle=-90]{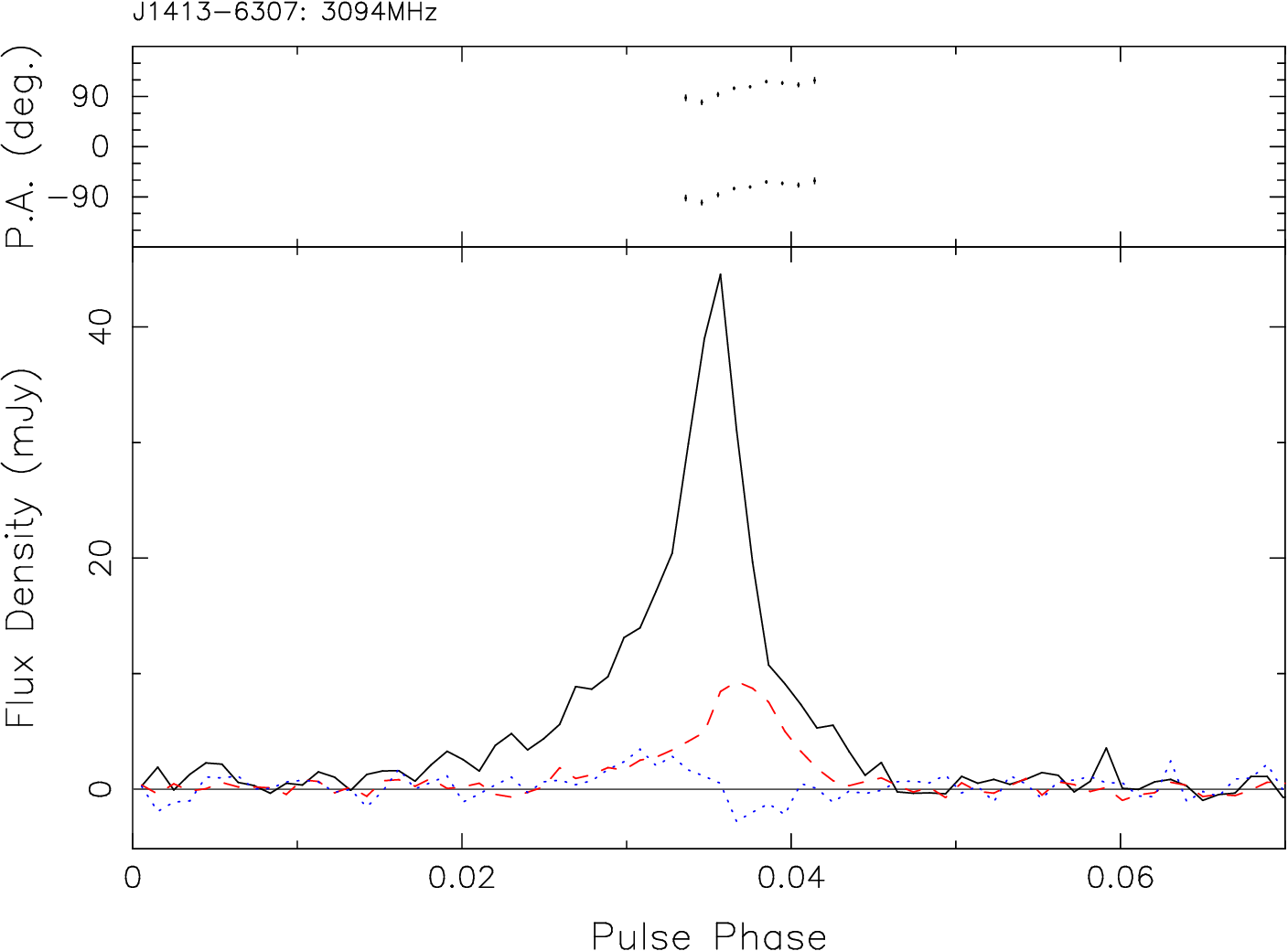}}& 
\resizebox{0.4\hsize}{!}{\includegraphics[angle=-90]{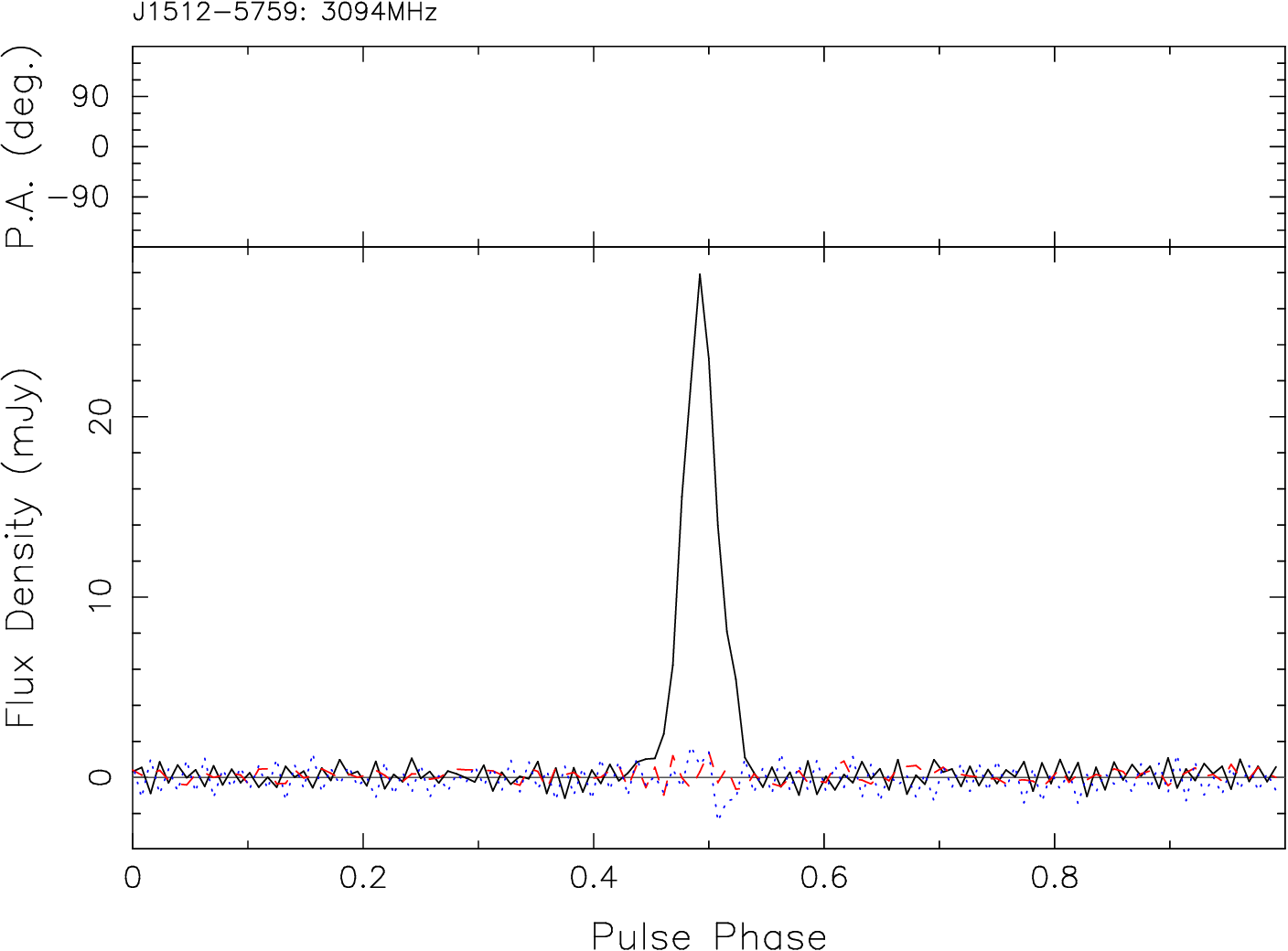}}\\
\end{tabular}
\end{center}
\caption{- continued.}
\end{figure*}
\addtocounter{figure}{-1}
\begin{figure*}
\begin{center}
\begin{tabular}{cc}
\resizebox{0.4\hsize}{!}{\includegraphics[angle=-90]{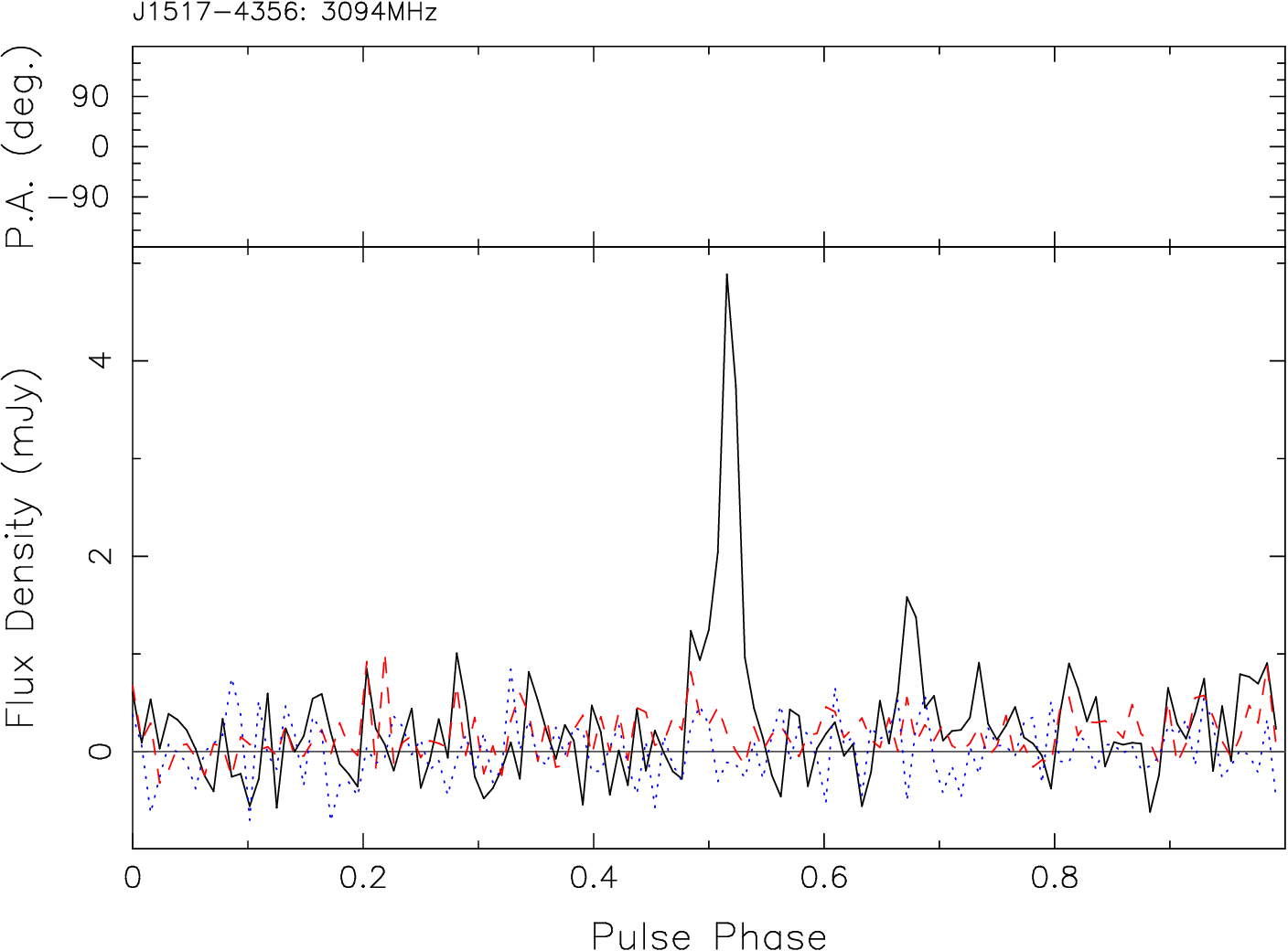}}& 
\resizebox{0.4\hsize}{!}{\includegraphics[angle=-90]{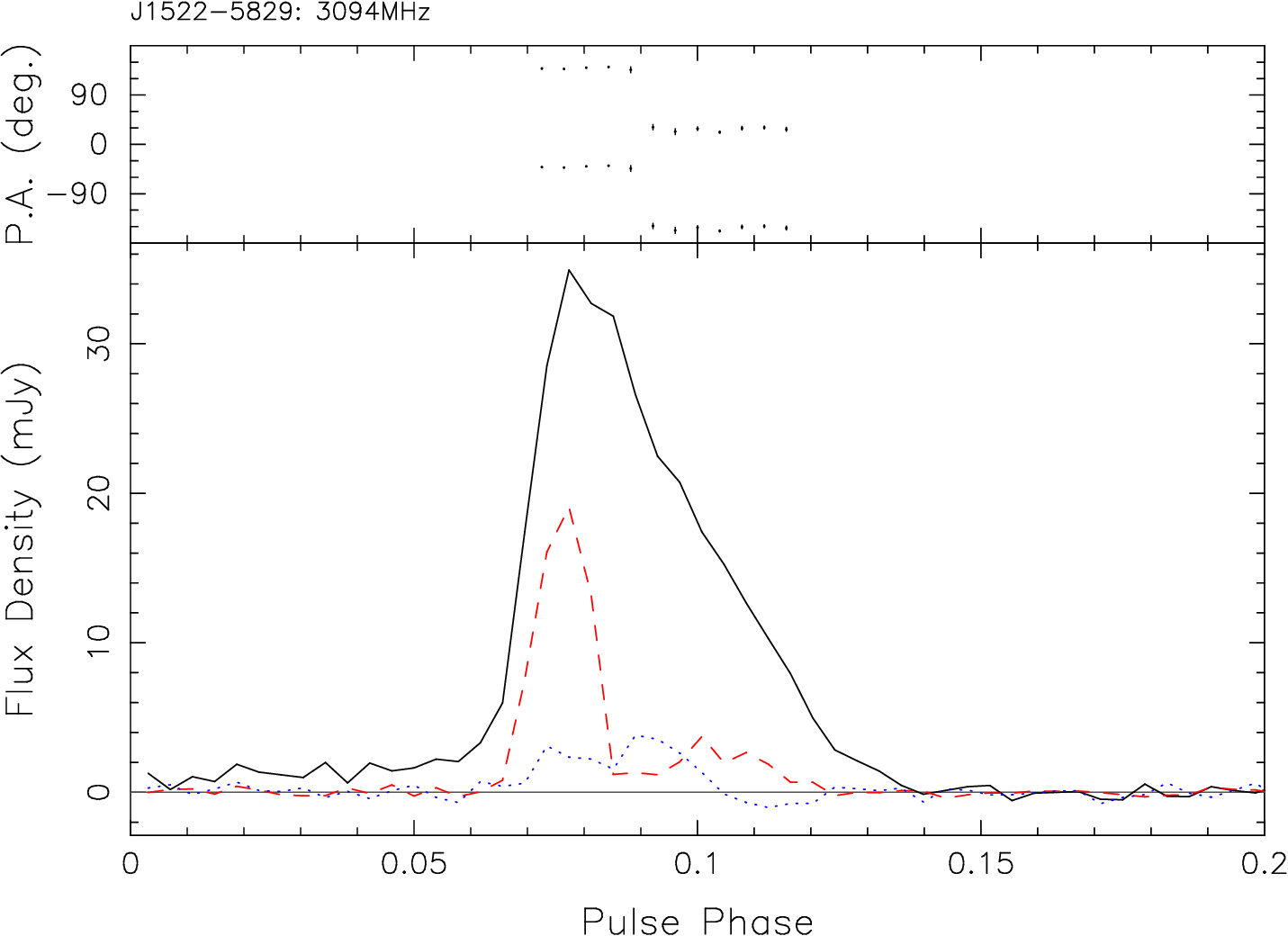}}\\
\resizebox{0.4\hsize}{!}{\includegraphics[angle=-90]{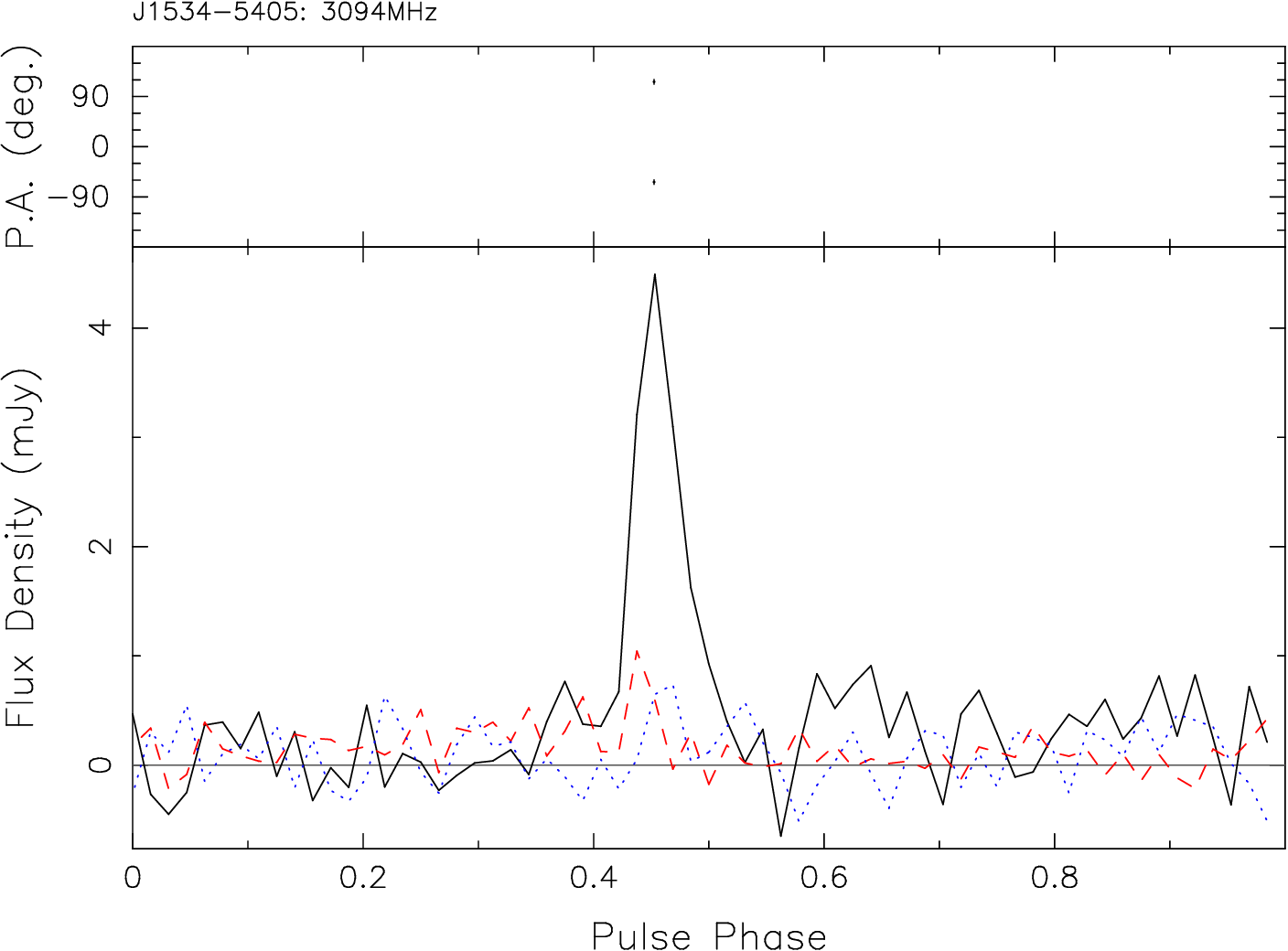}}& 
\resizebox{0.4\hsize}{!}{\includegraphics[angle=-90]{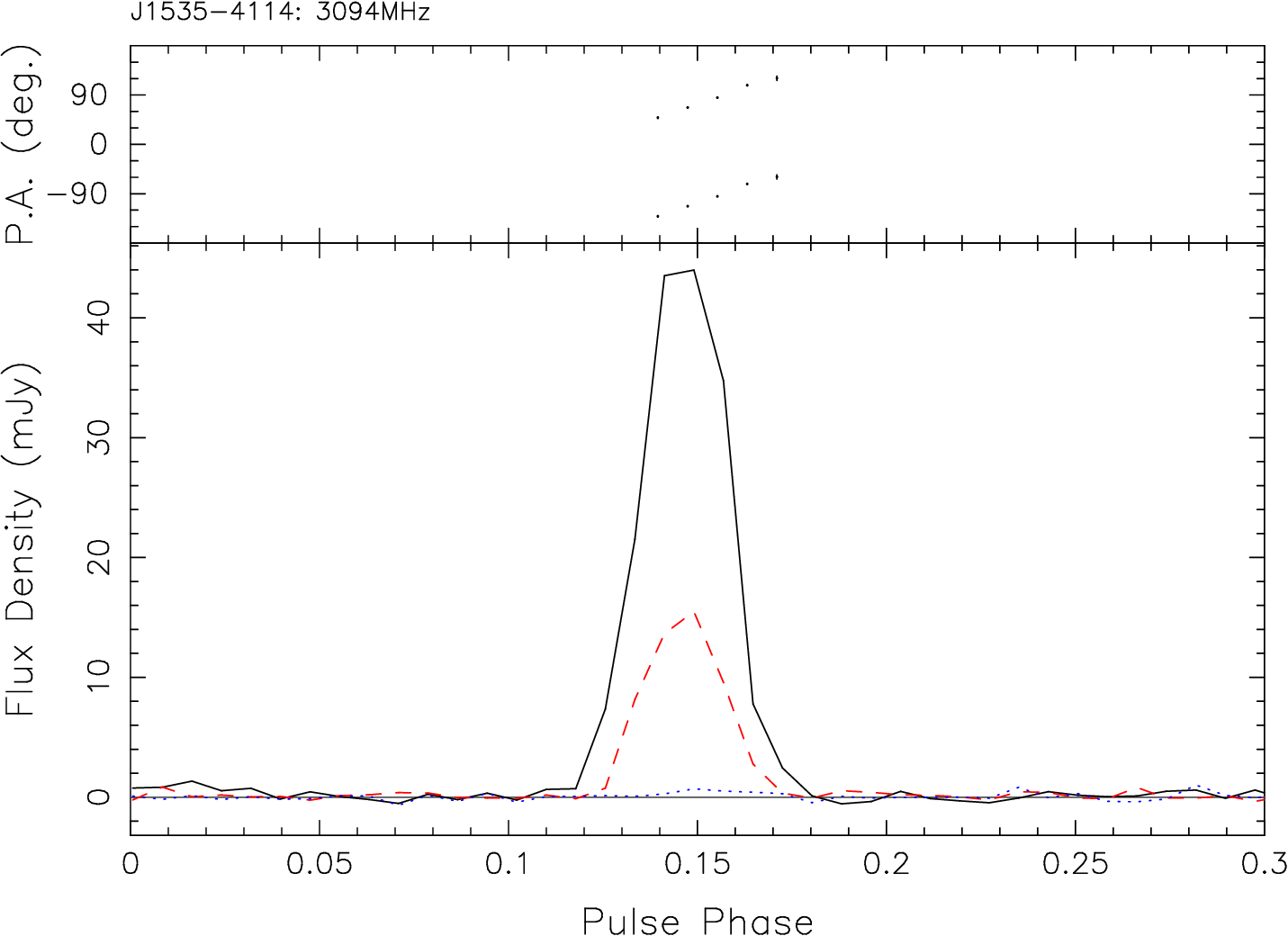}}\\
\resizebox{0.4\hsize}{!}{\includegraphics[angle=-90]{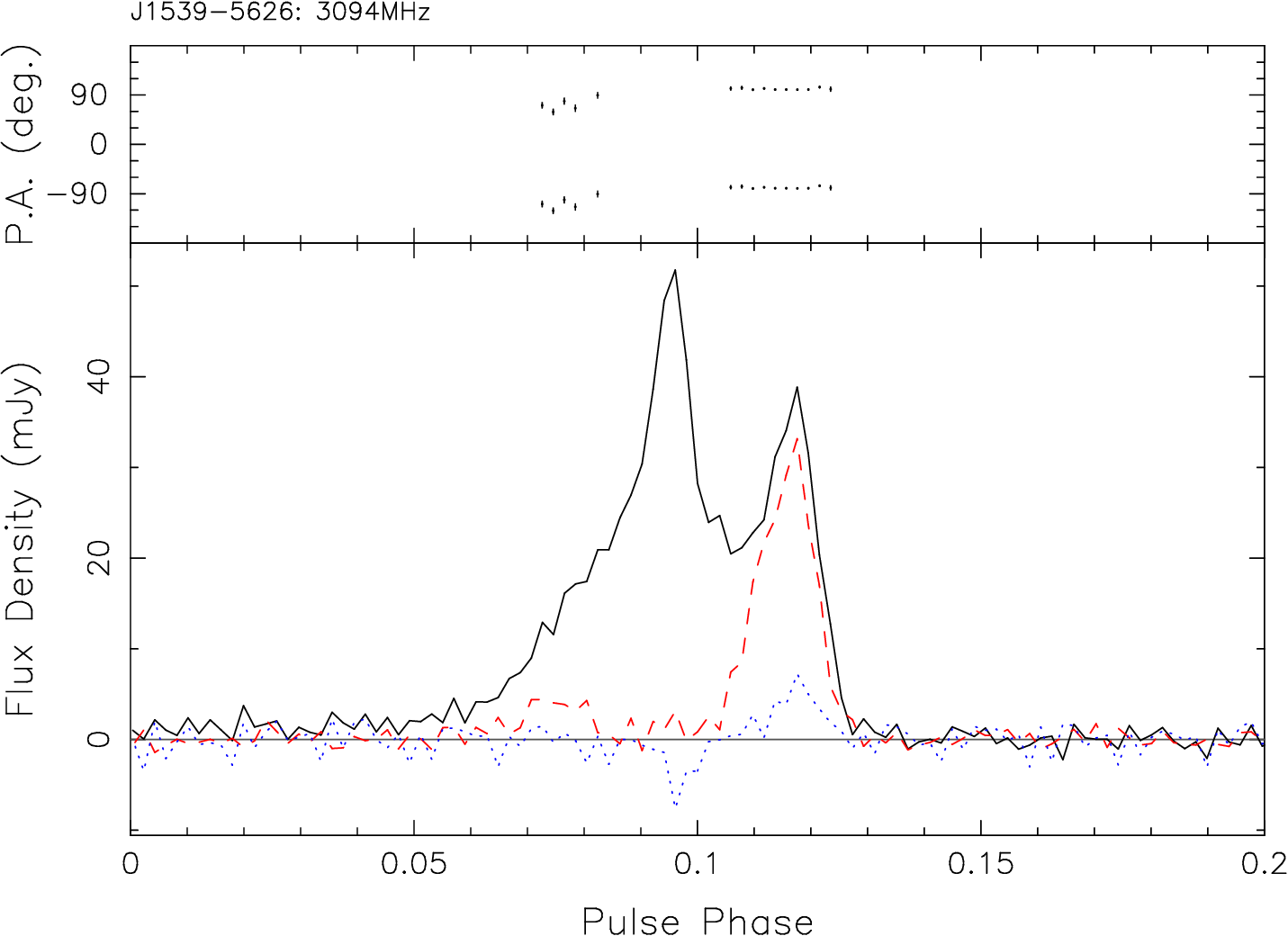}}& 
\resizebox{0.4\hsize}{!}{\includegraphics[angle=-90]{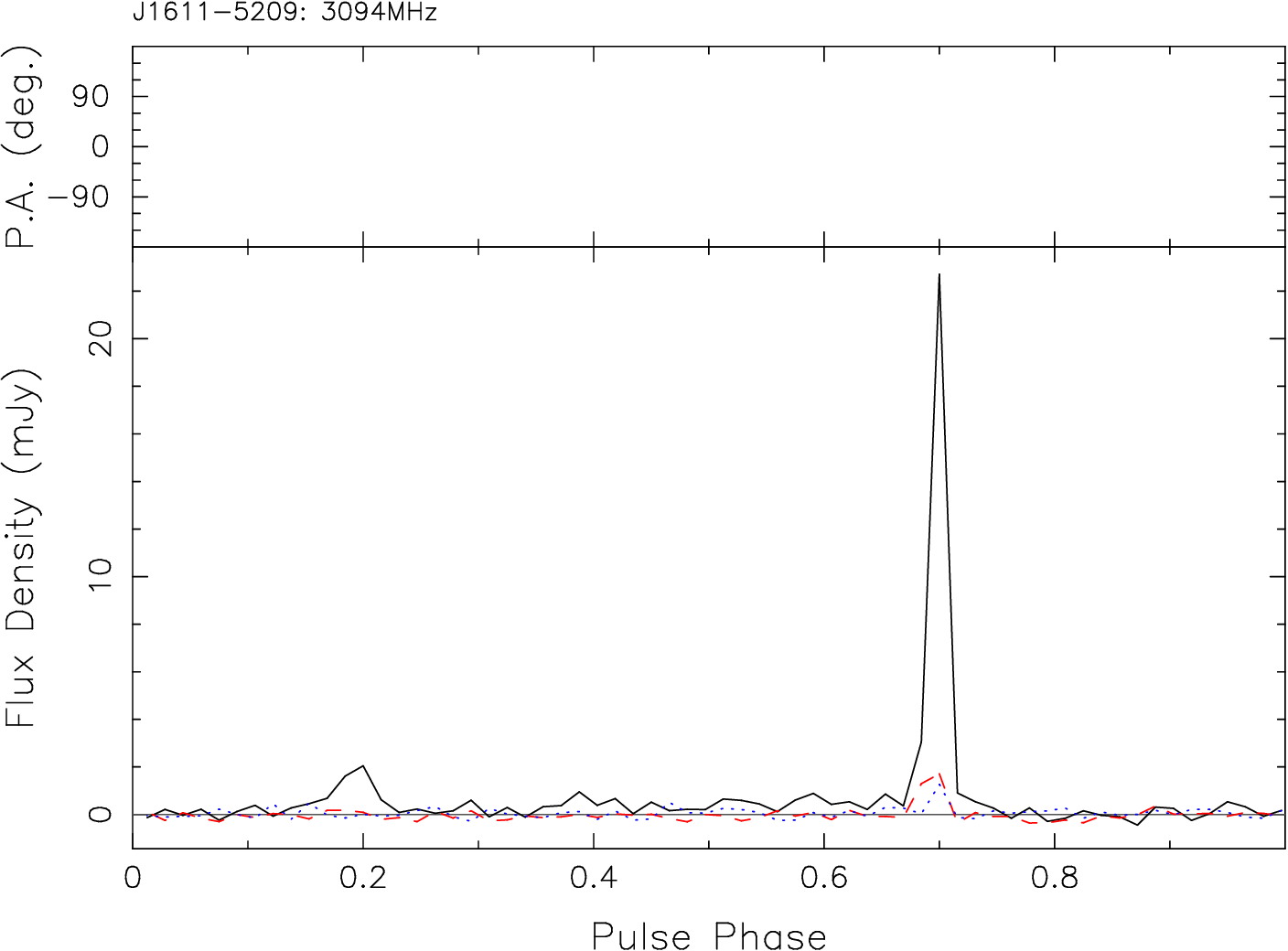}}\\
\resizebox{0.4\hsize}{!}{\includegraphics[angle=-90]{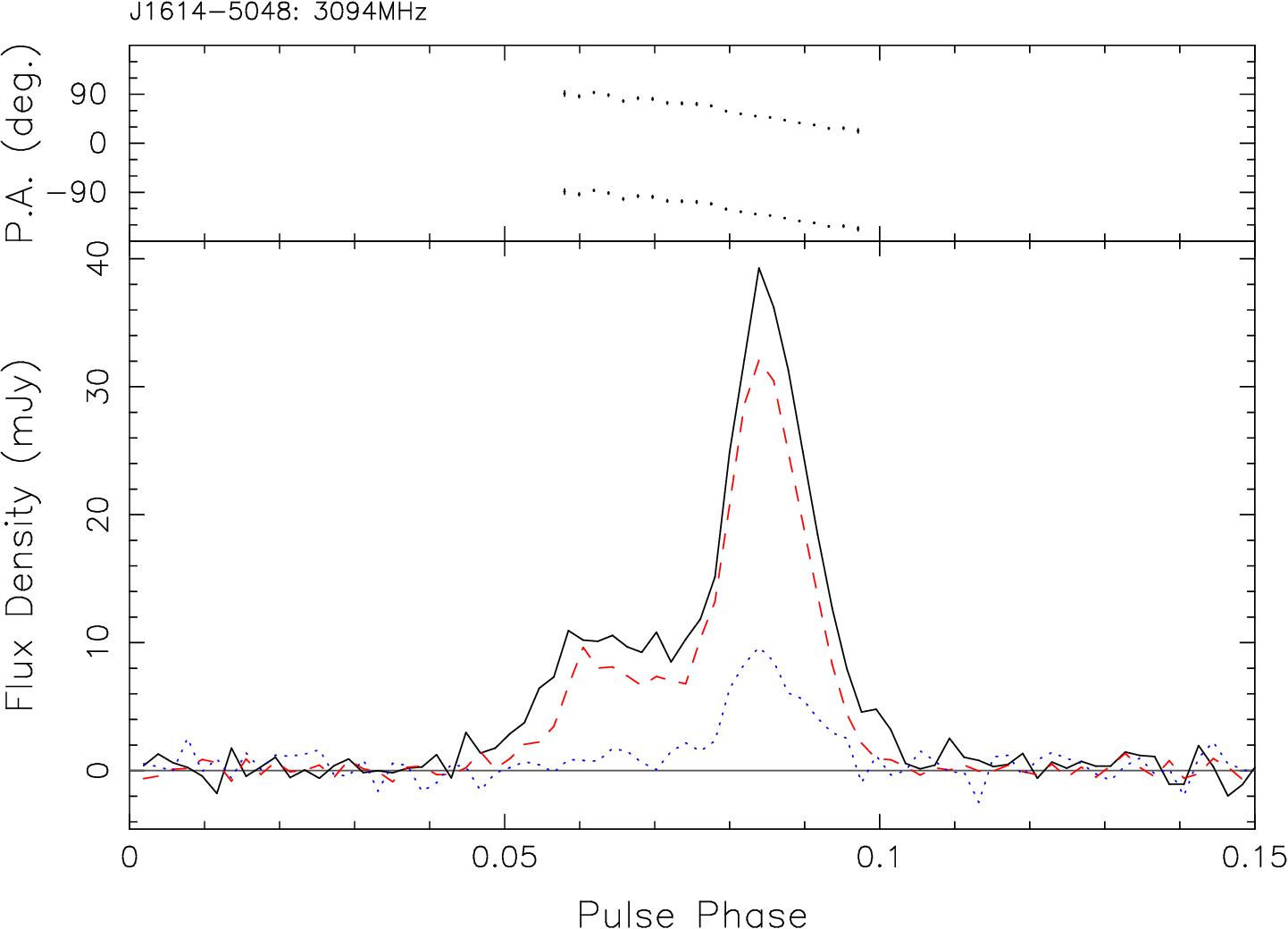}}&  
\resizebox{0.4\hsize}{!}{\includegraphics[angle=-90]{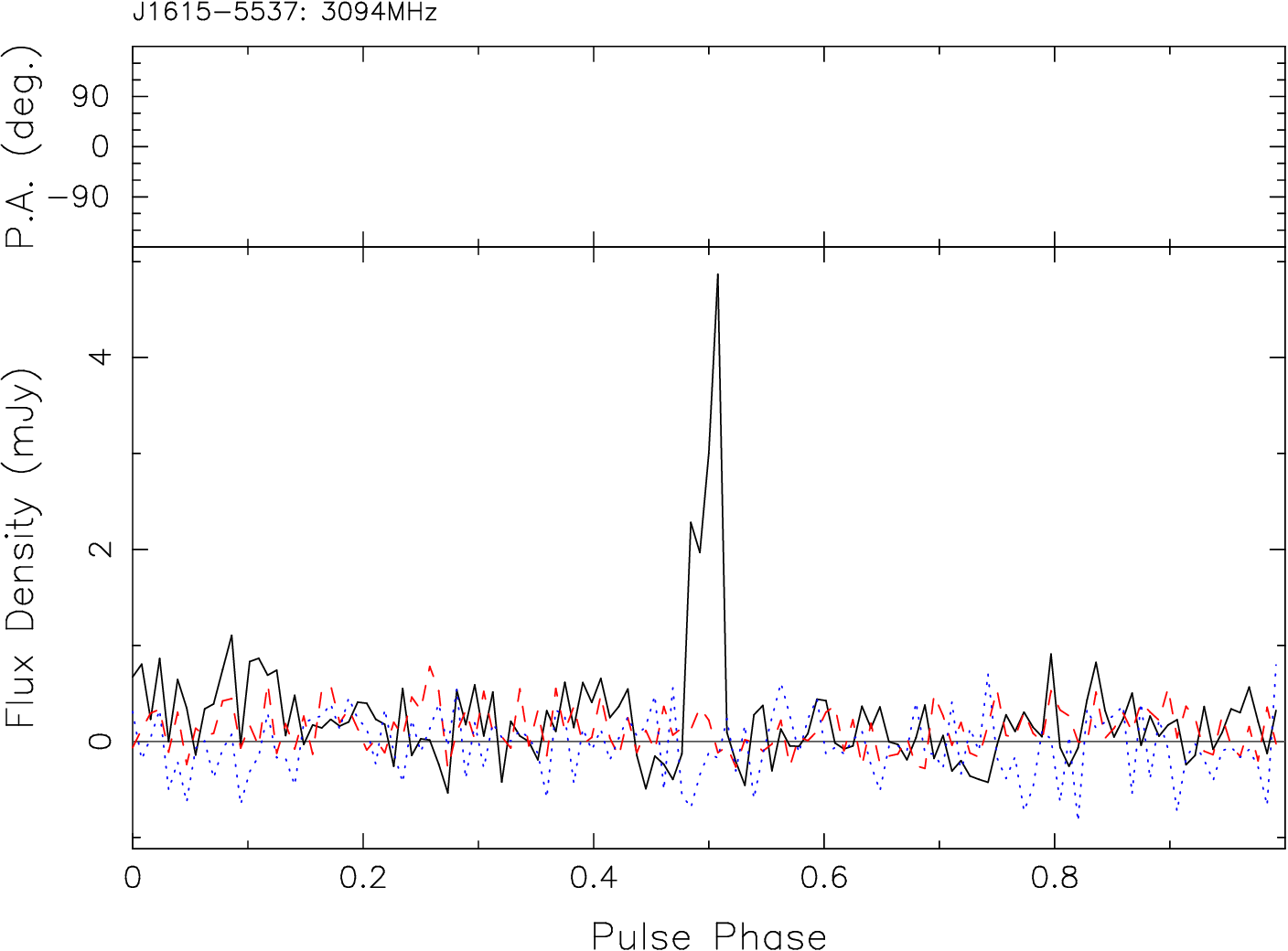}}\\
\end{tabular}
\end{center}
\caption{- continued.}
\end{figure*}
\addtocounter{figure}{-1}
\begin{figure*}
\begin{center}
\begin{tabular}{cc}
\resizebox{0.4\hsize}{!}{\includegraphics[angle=-90]{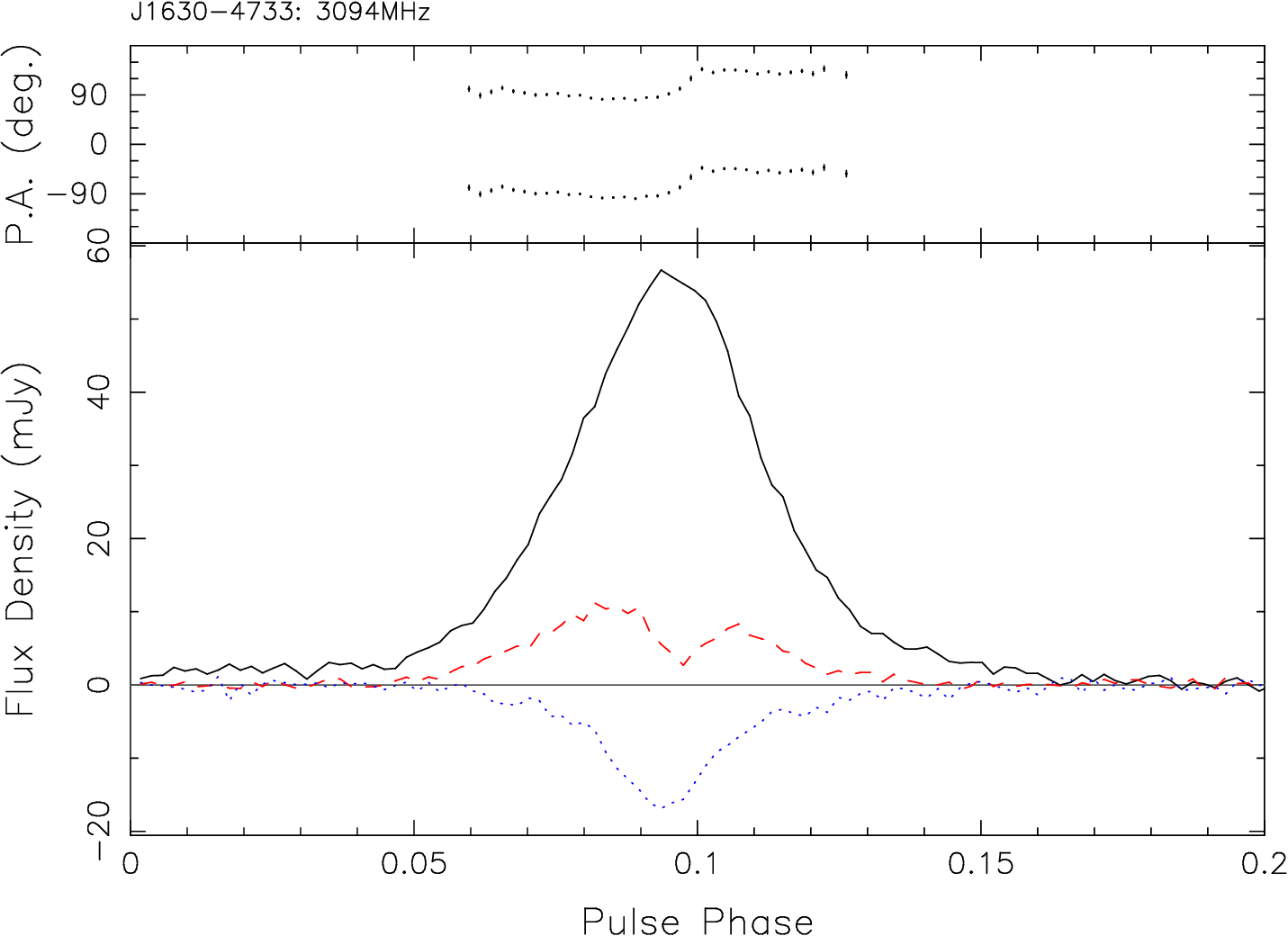}}& 
\resizebox{0.4\hsize}{!}{\includegraphics[angle=-90]{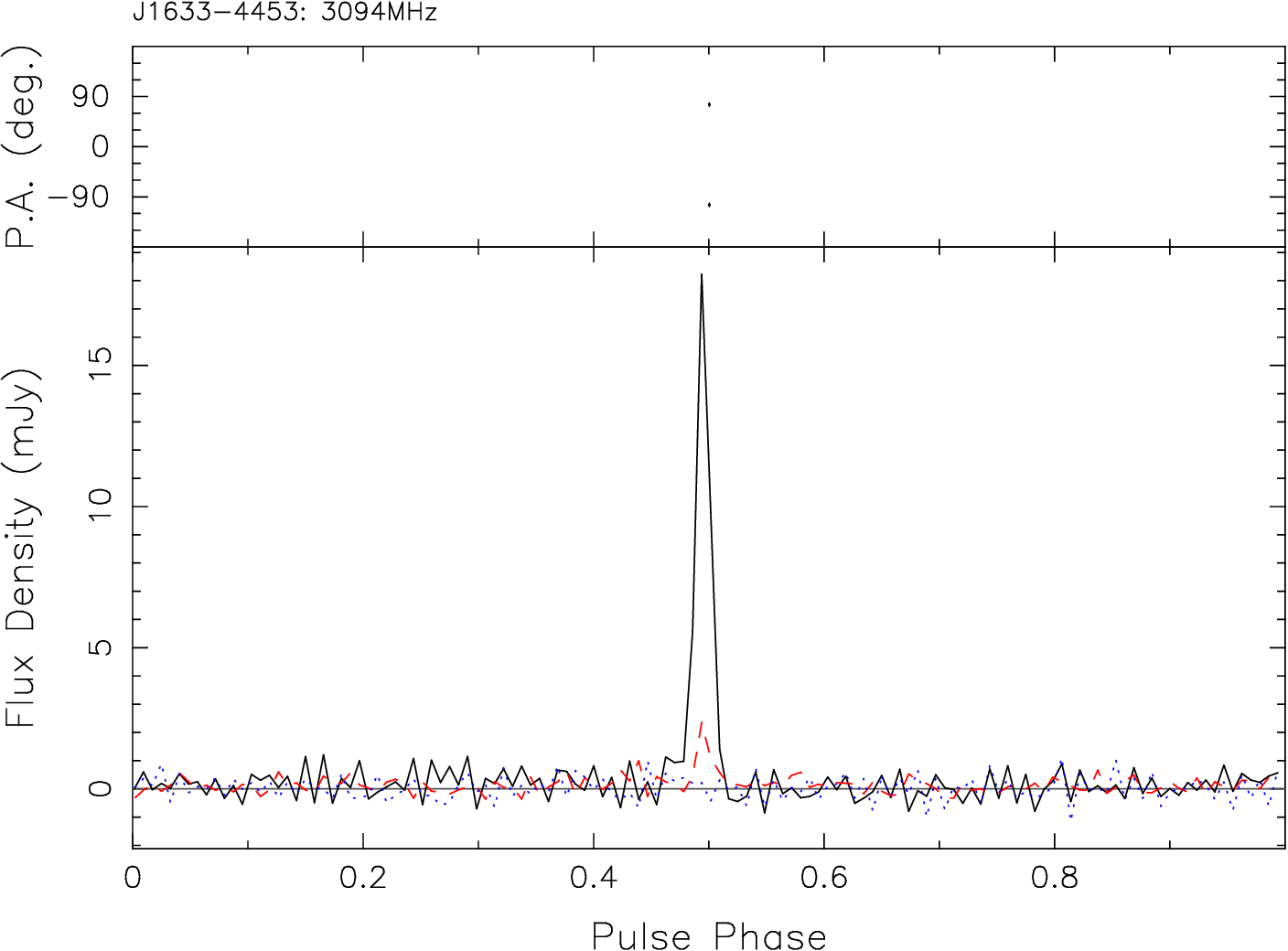}}\\
\resizebox{0.4\hsize}{!}{\includegraphics[angle=-90]{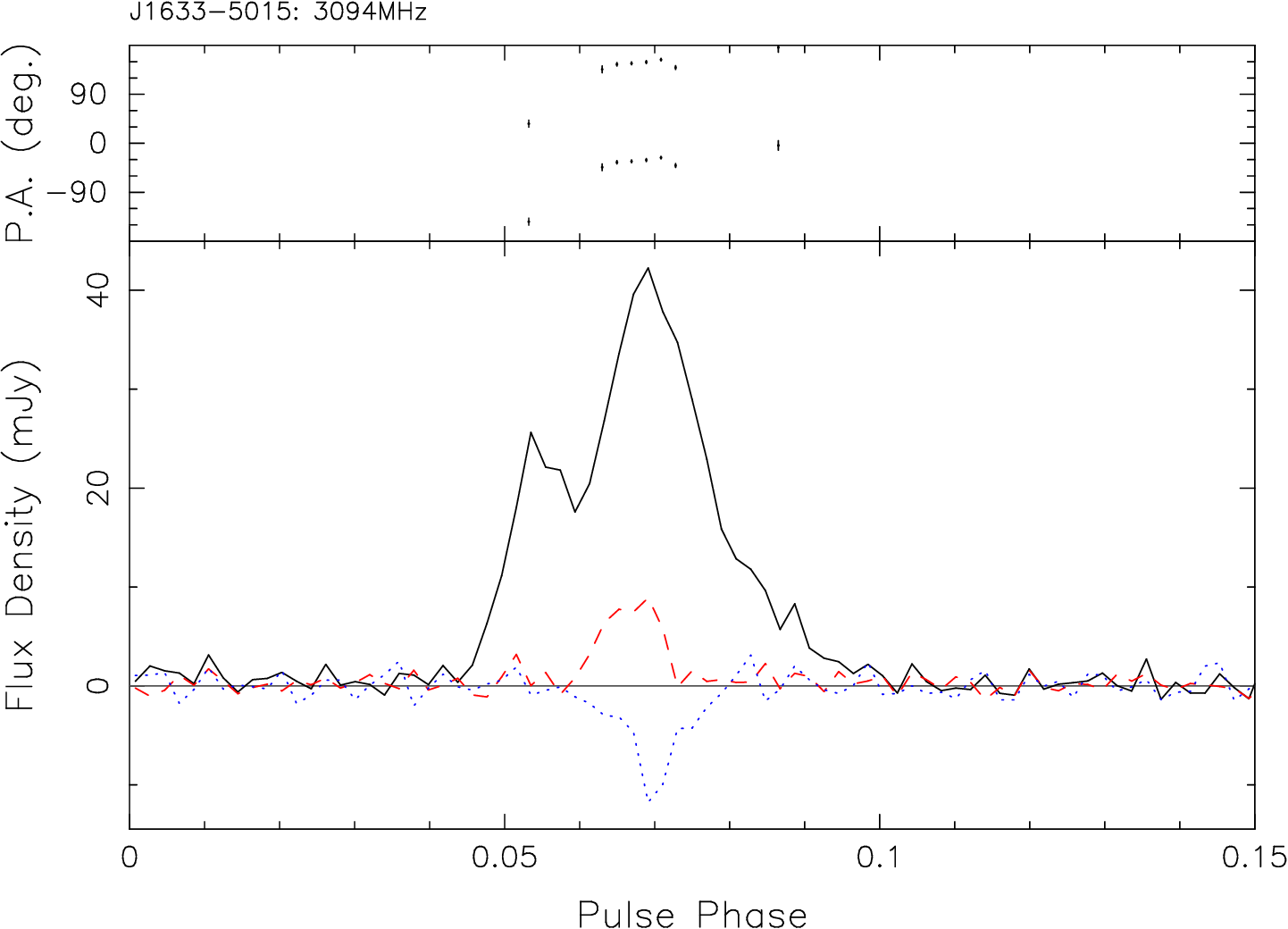}}& 
\resizebox{0.4\hsize}{!}{\includegraphics[angle=-90]{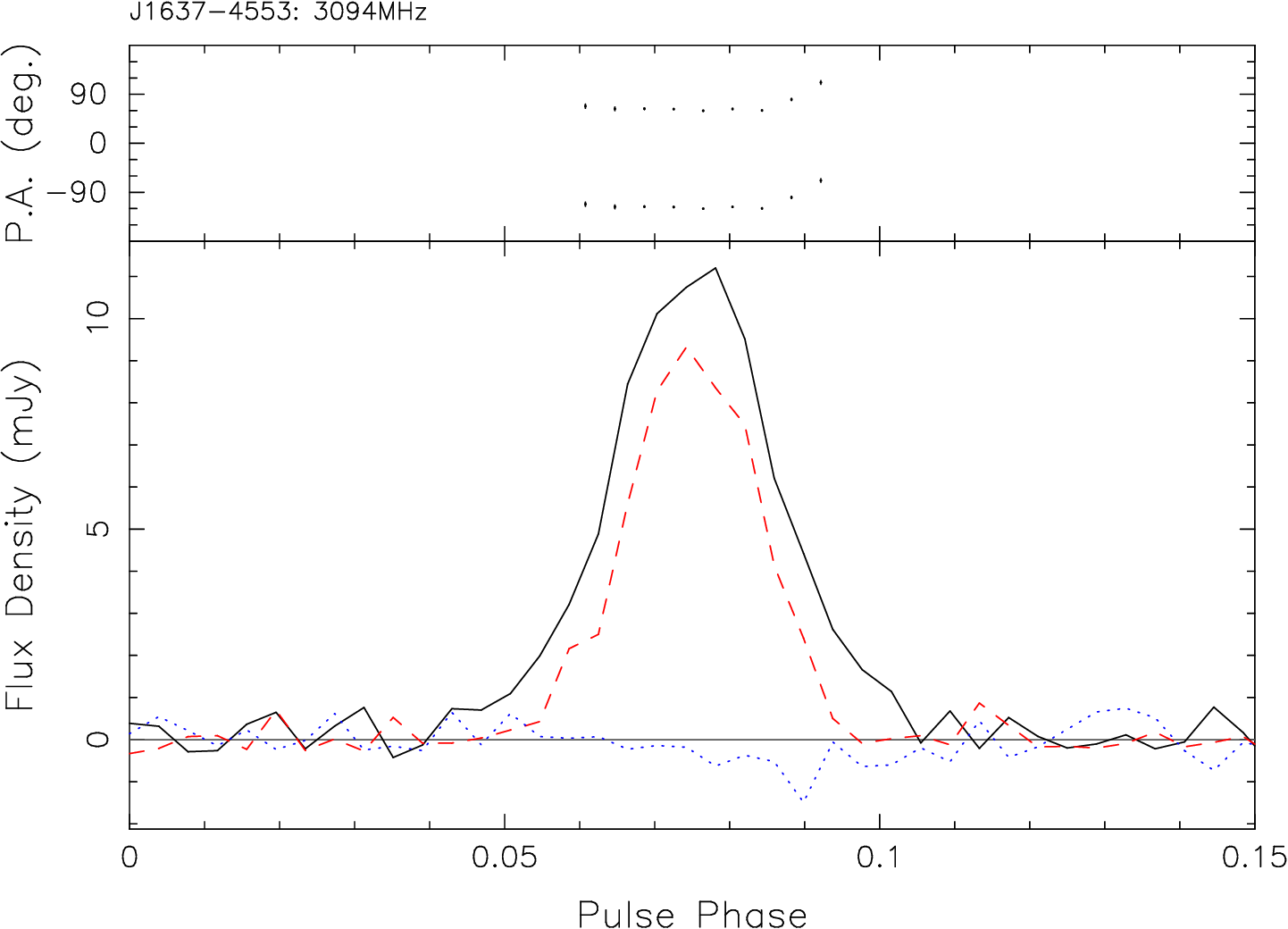}}\\
\resizebox{0.4\hsize}{!}{\includegraphics[angle=-90]{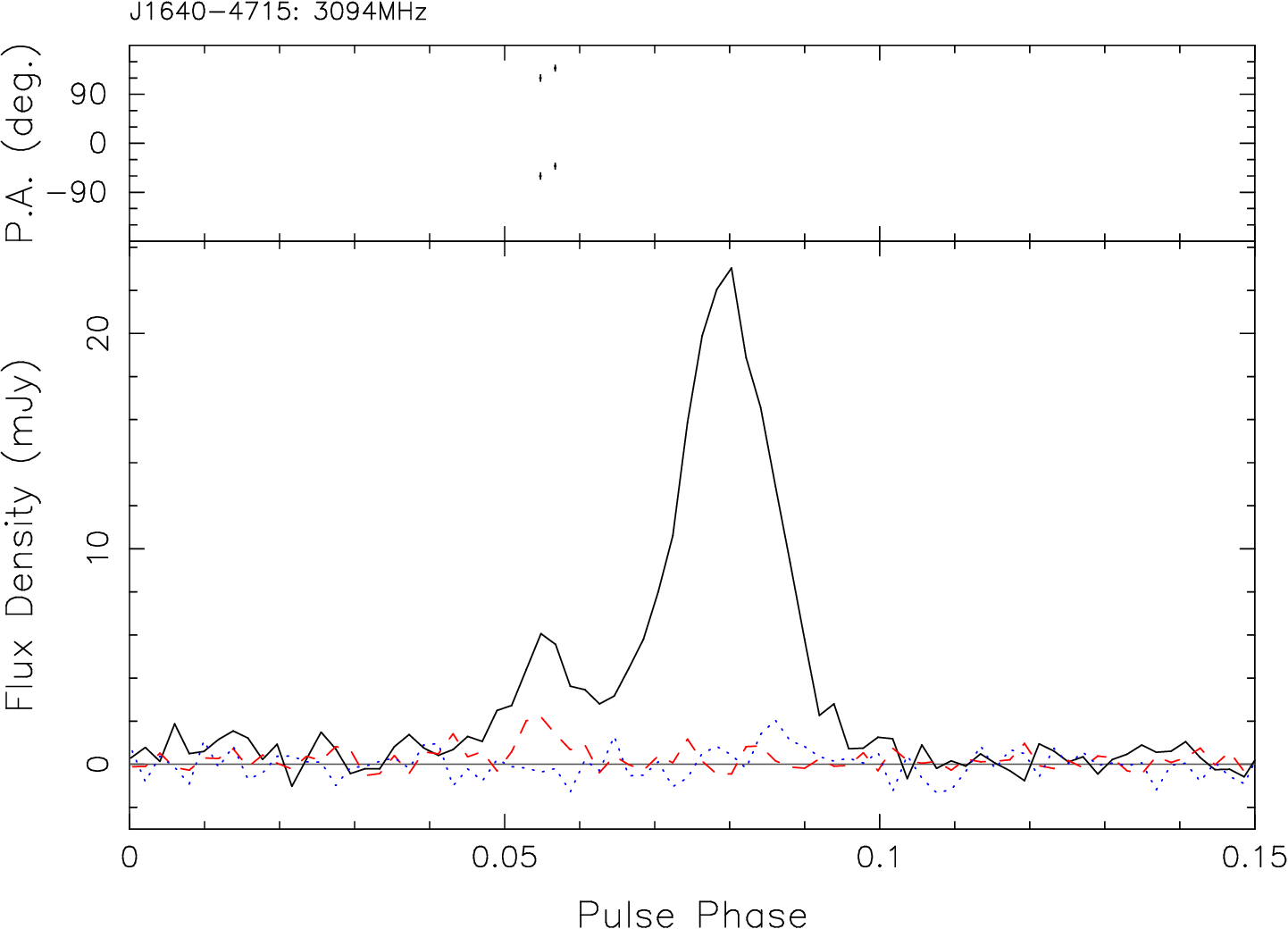}}& 
\resizebox{0.4\hsize}{!}{\includegraphics[angle=-90]{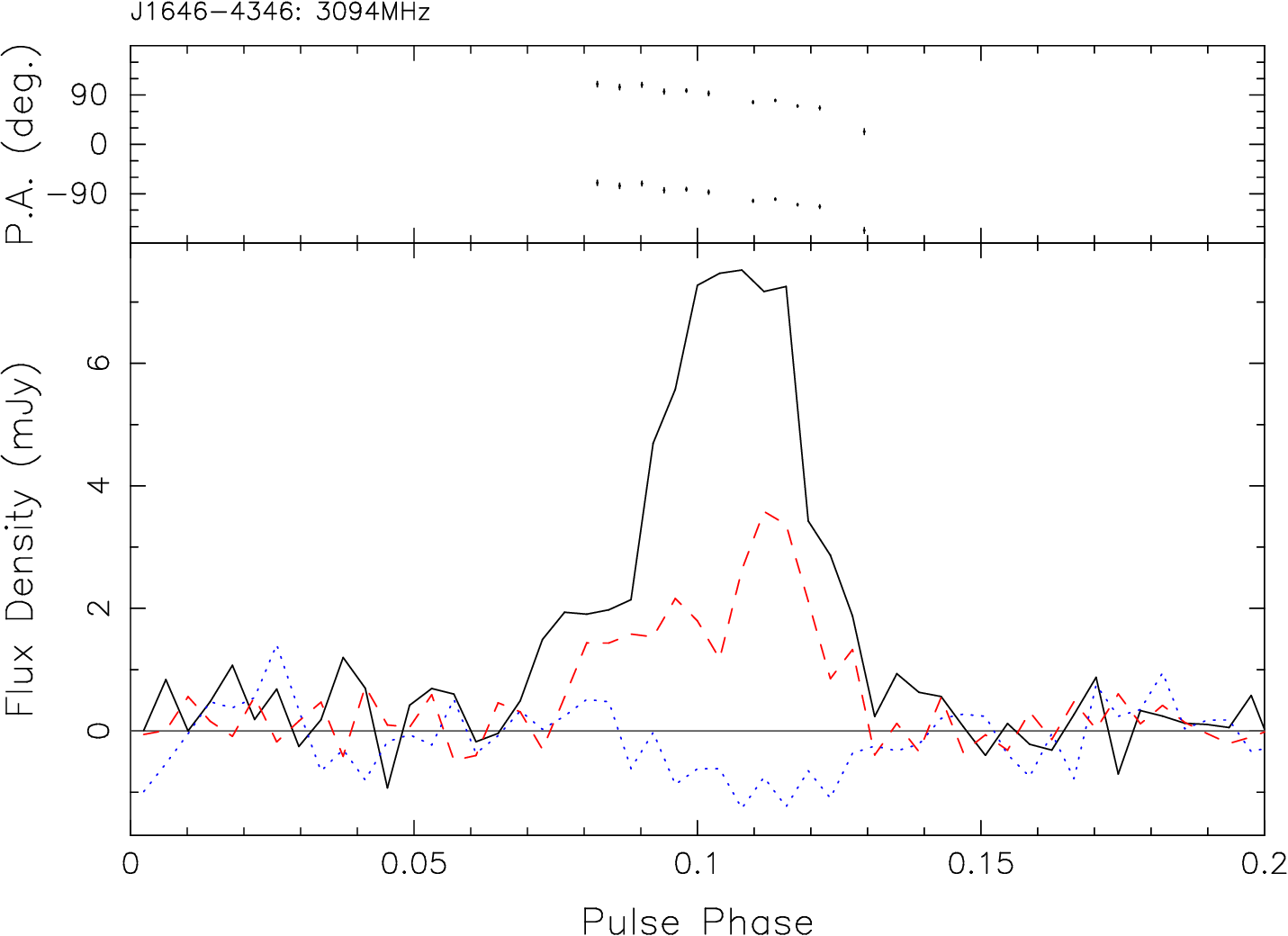}}\\
\resizebox{0.4\hsize}{!}{\includegraphics[angle=-90]{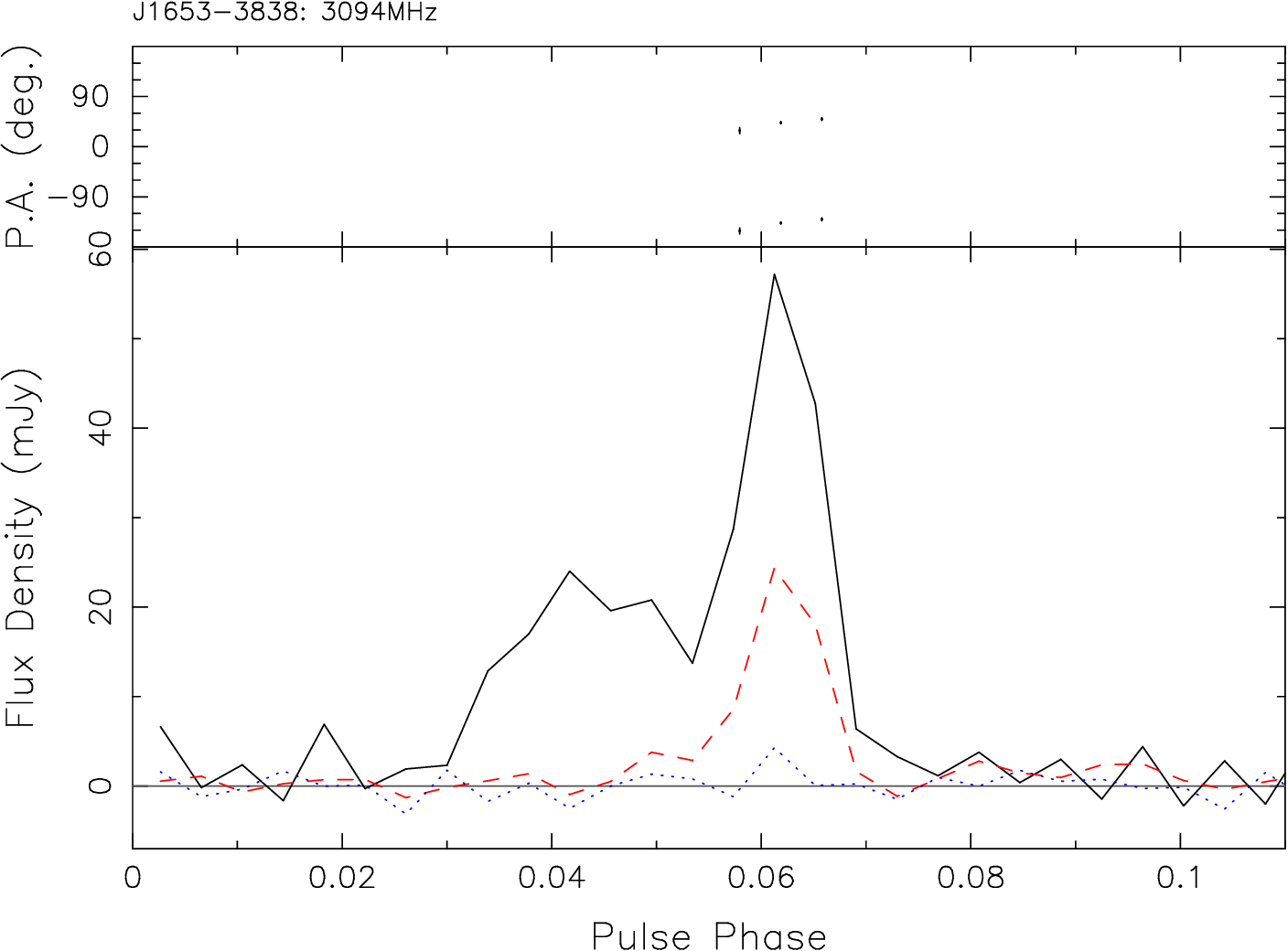}}& 
\resizebox{0.4\hsize}{!}{\includegraphics[angle=-90]{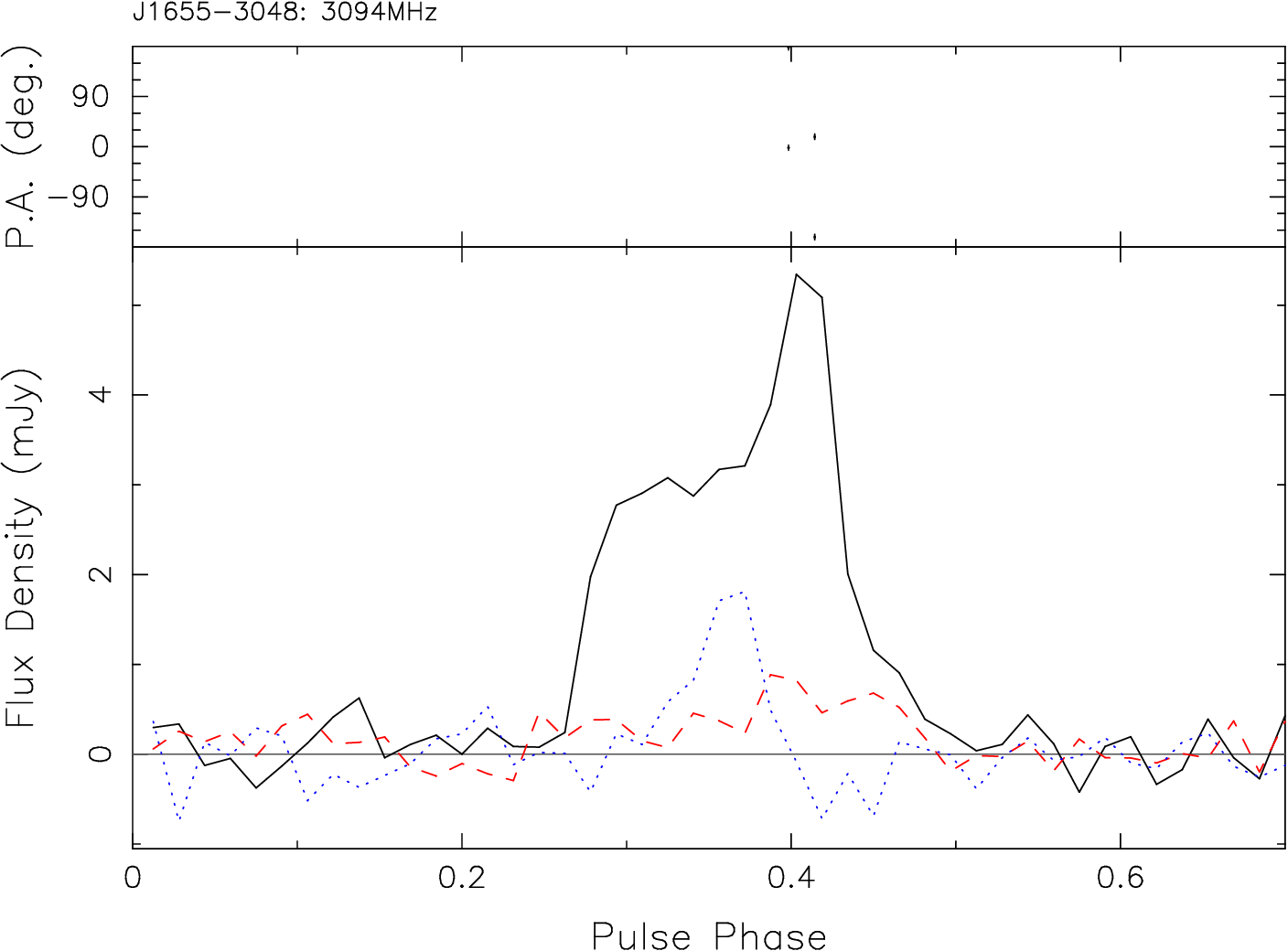}}\\
\end{tabular}
\end{center}
\caption{- continued.}
\end{figure*}
\addtocounter{figure}{-1}
\begin{figure*}
\begin{center}
\begin{tabular}{cc}
\resizebox{0.4\hsize}{!}{\includegraphics[angle=-90]{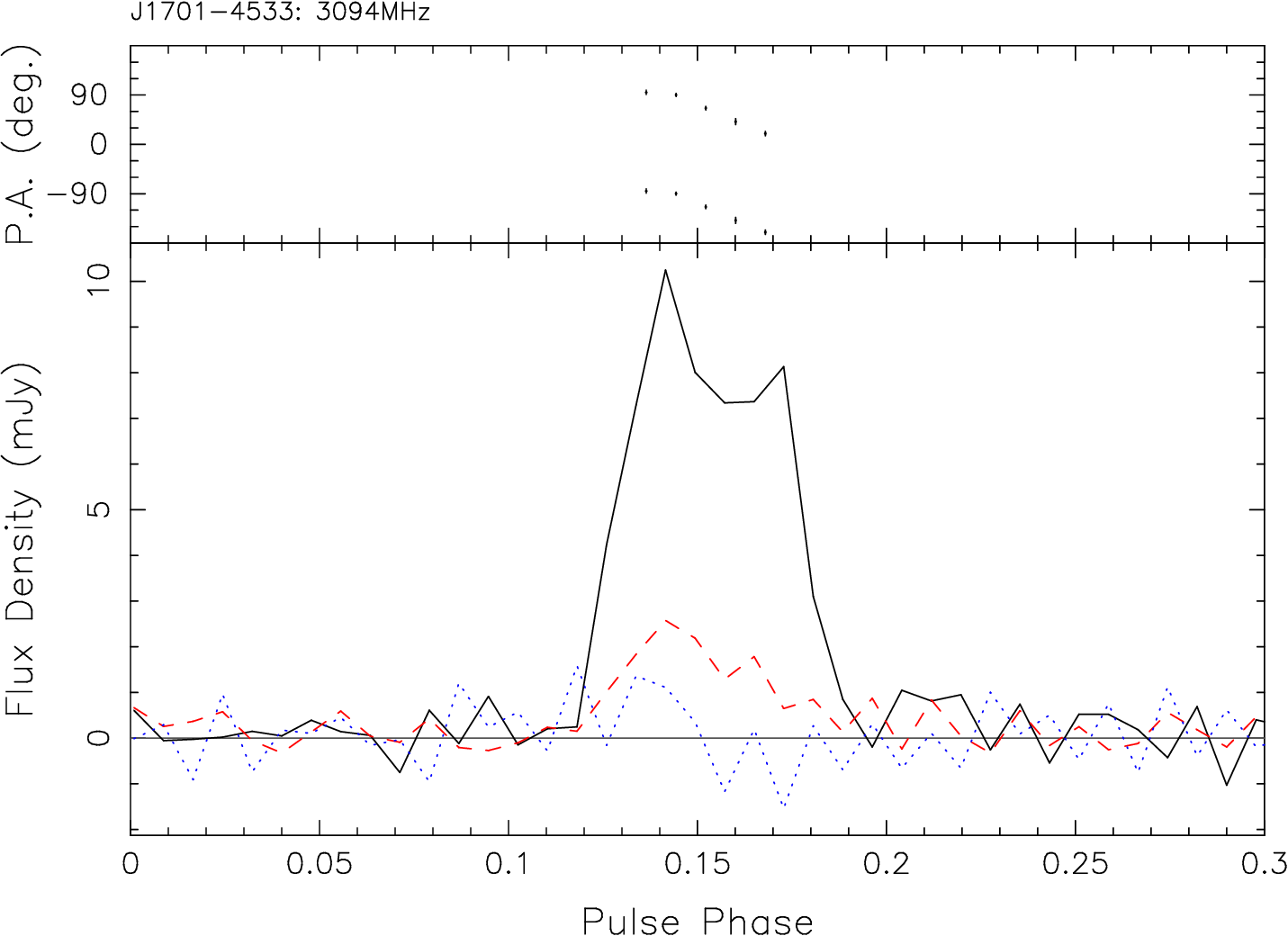}}& 
\resizebox{0.4\hsize}{!}{\includegraphics[angle=-90]{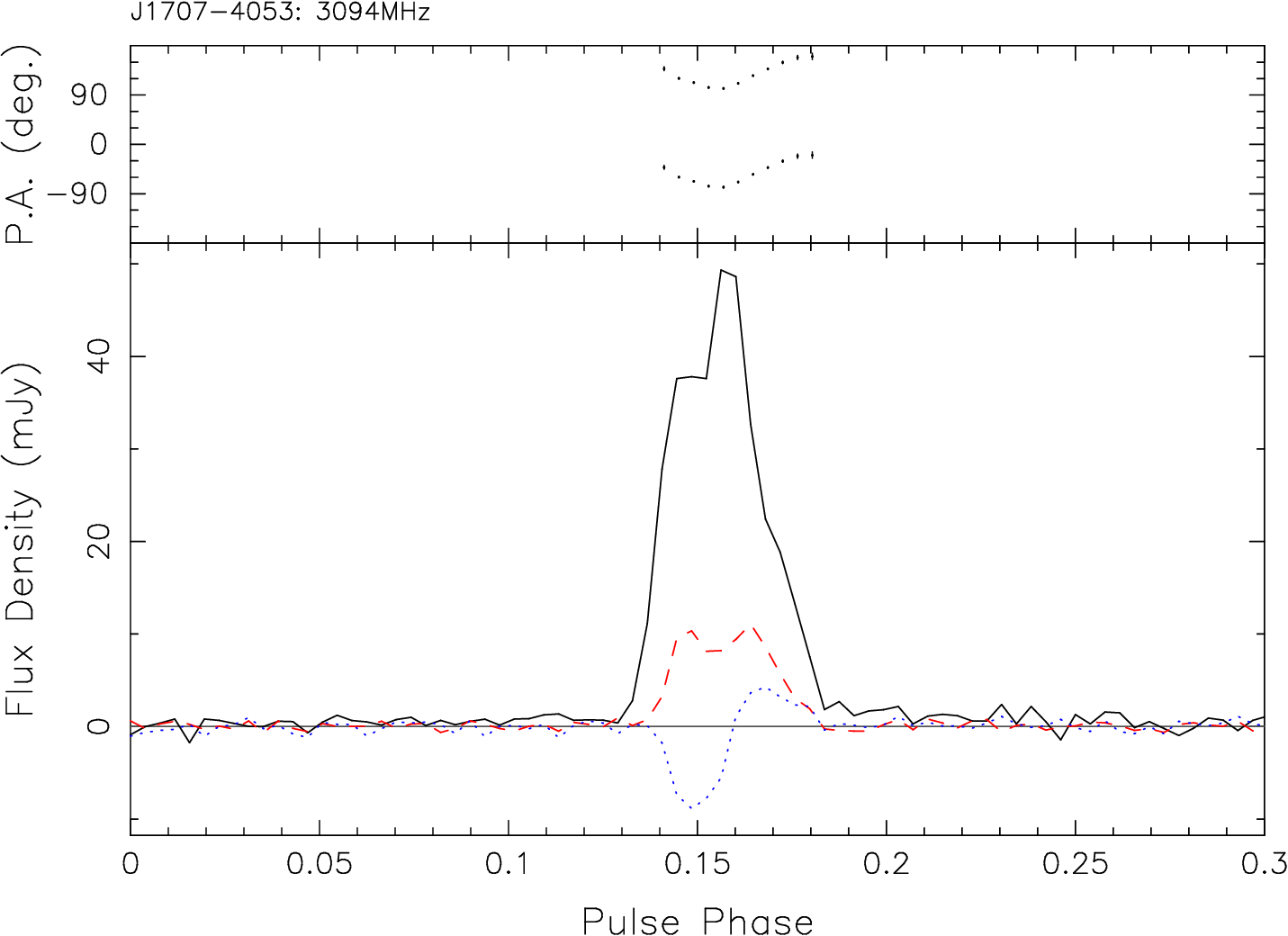}}\\
\resizebox{0.4\hsize}{!}{\includegraphics[angle=-90]{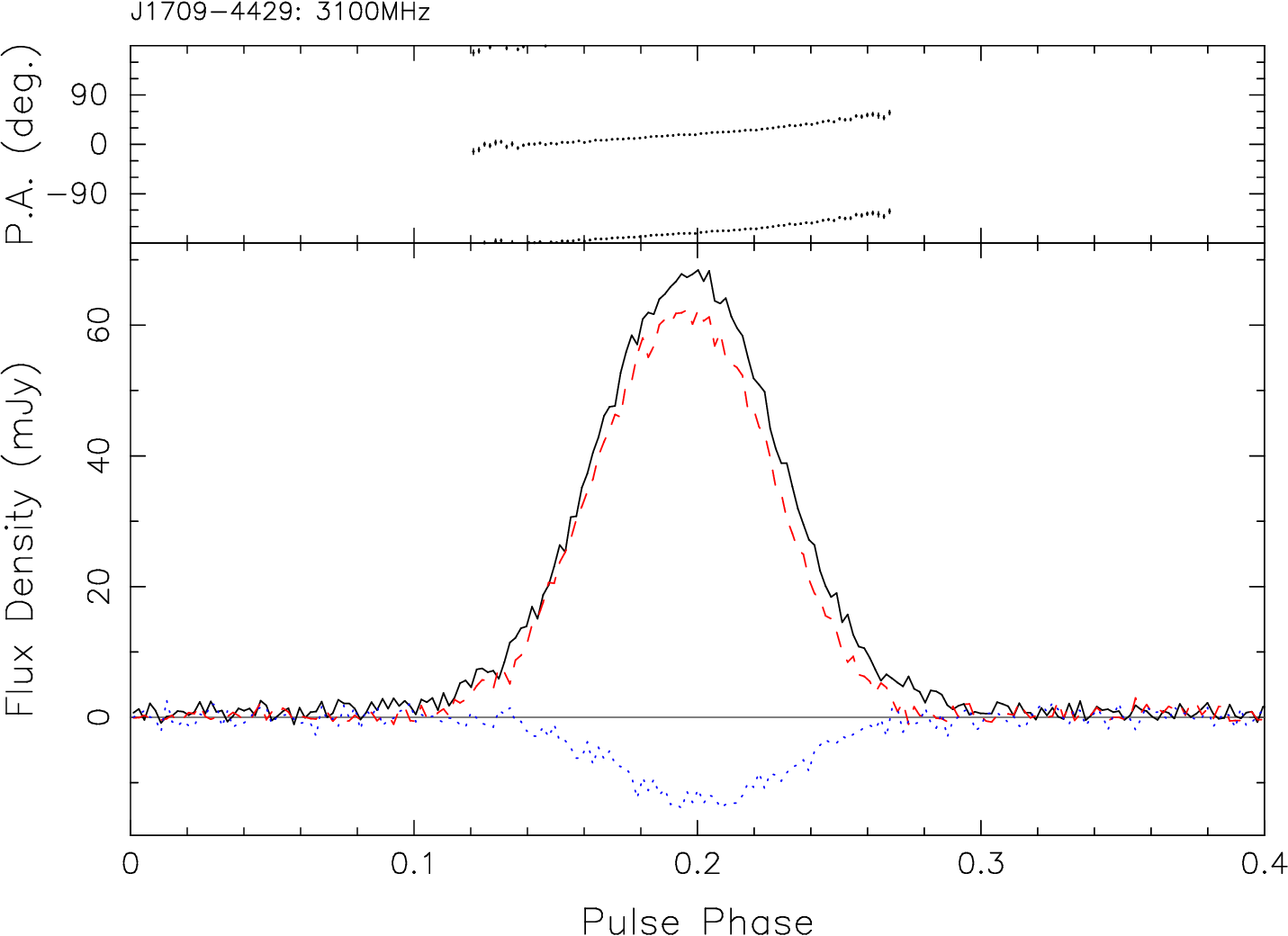}}& 
\resizebox{0.4\hsize}{!}{\includegraphics[angle=-90]{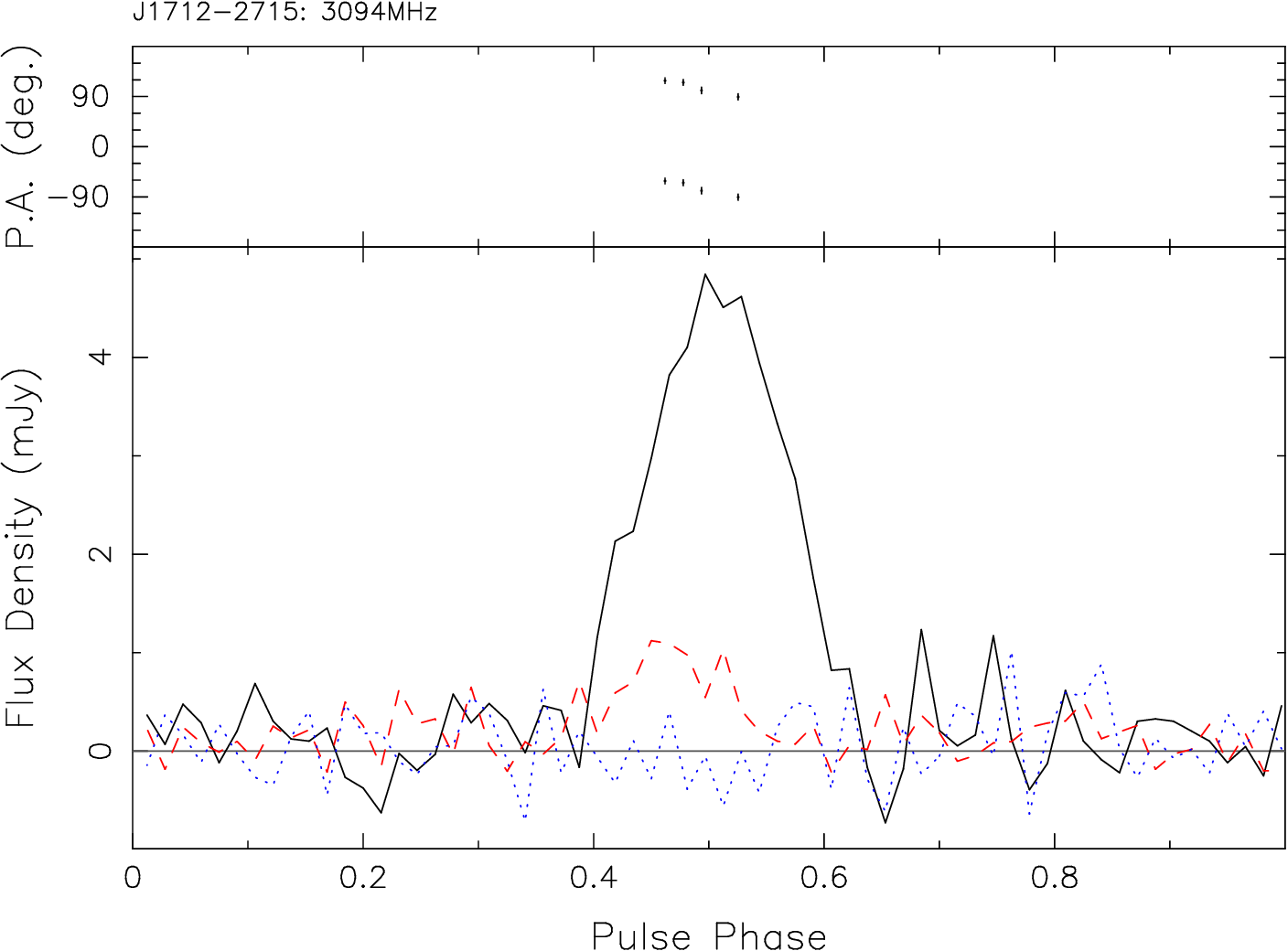}}\\
\resizebox{0.4\hsize}{!}{\includegraphics[angle=-90]{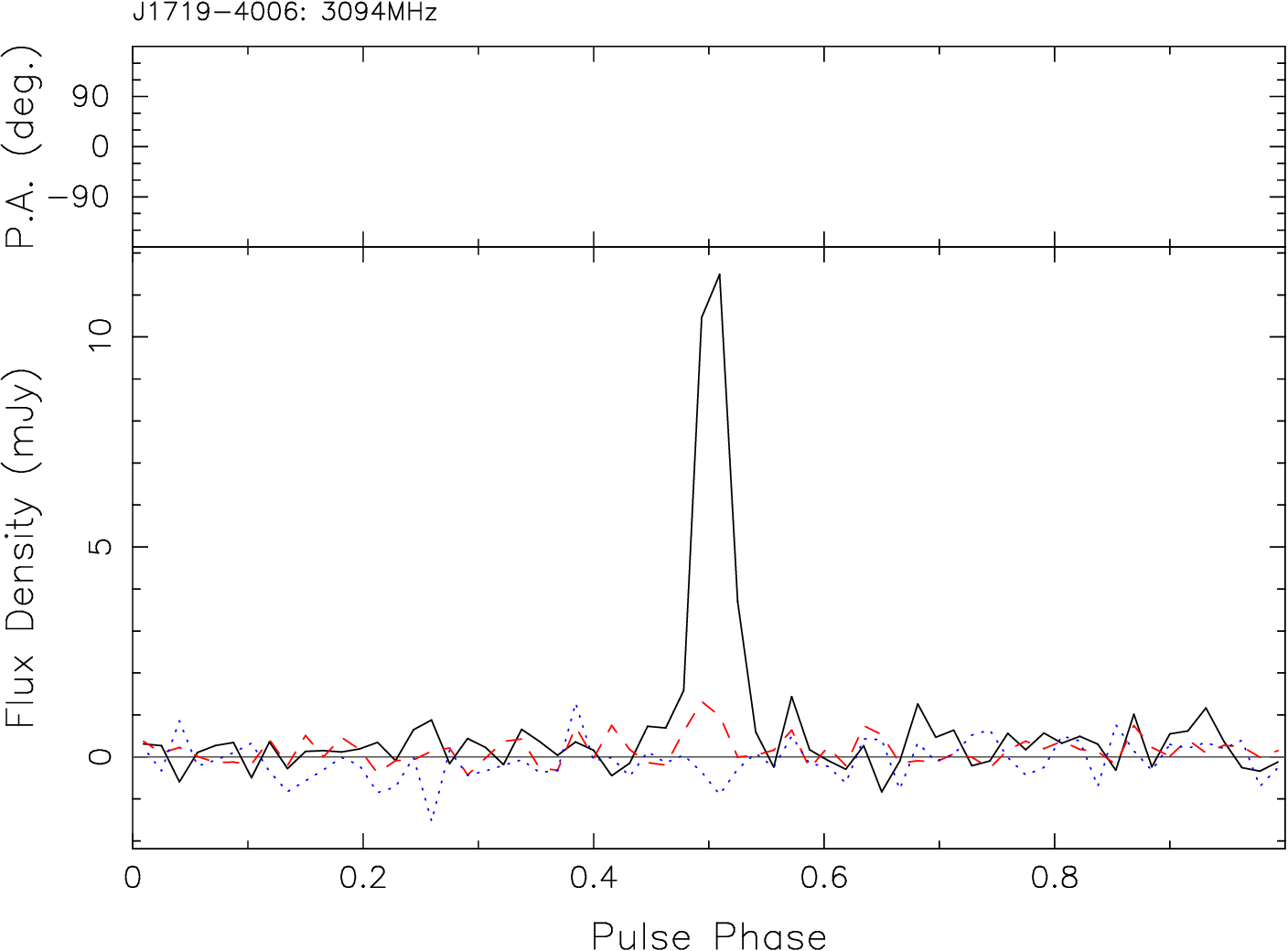}}& 
\resizebox{0.4\hsize}{!}{\includegraphics[angle=-90]{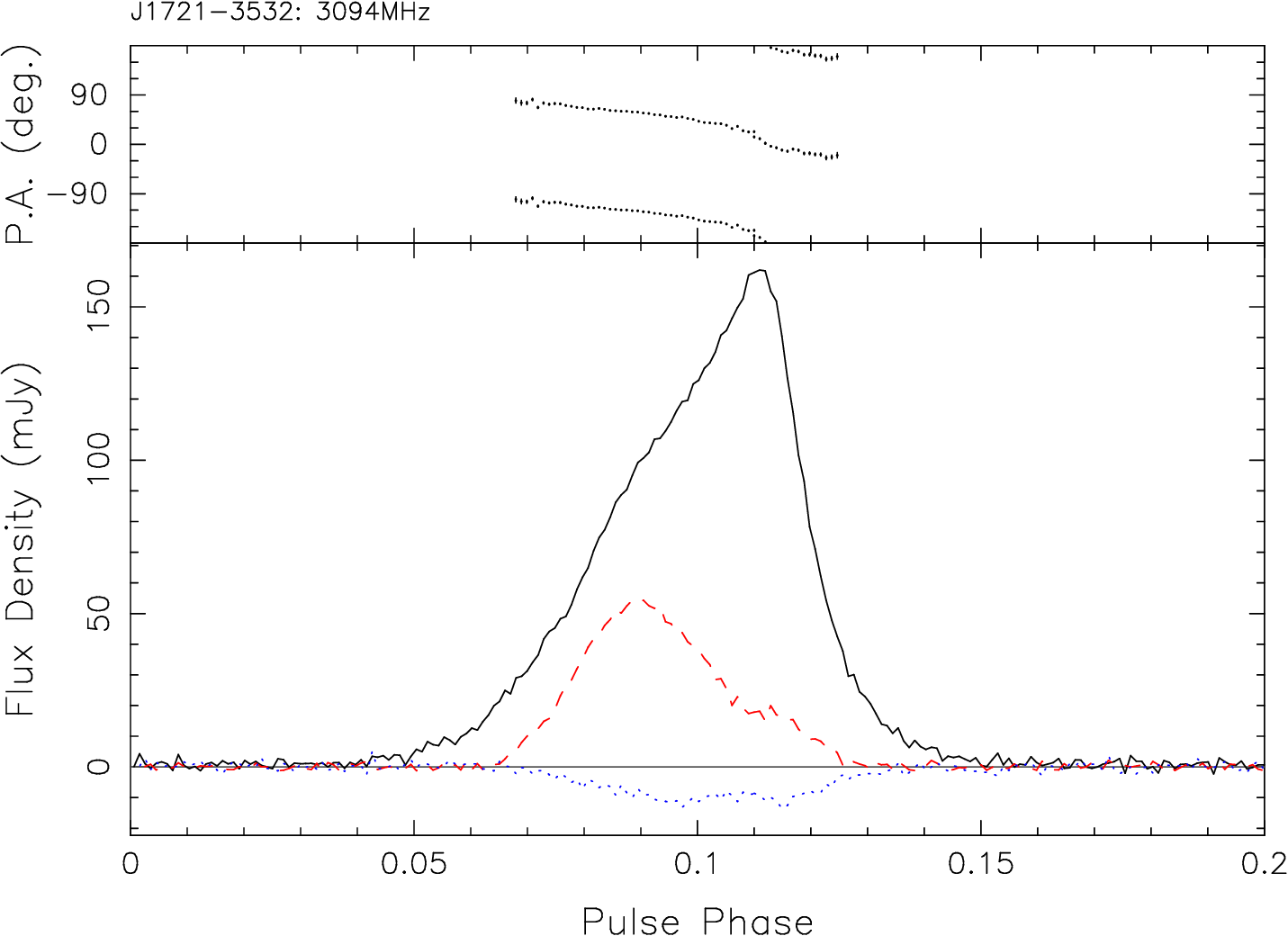}}\\
\resizebox{0.4\hsize}{!}{\includegraphics[angle=-90]{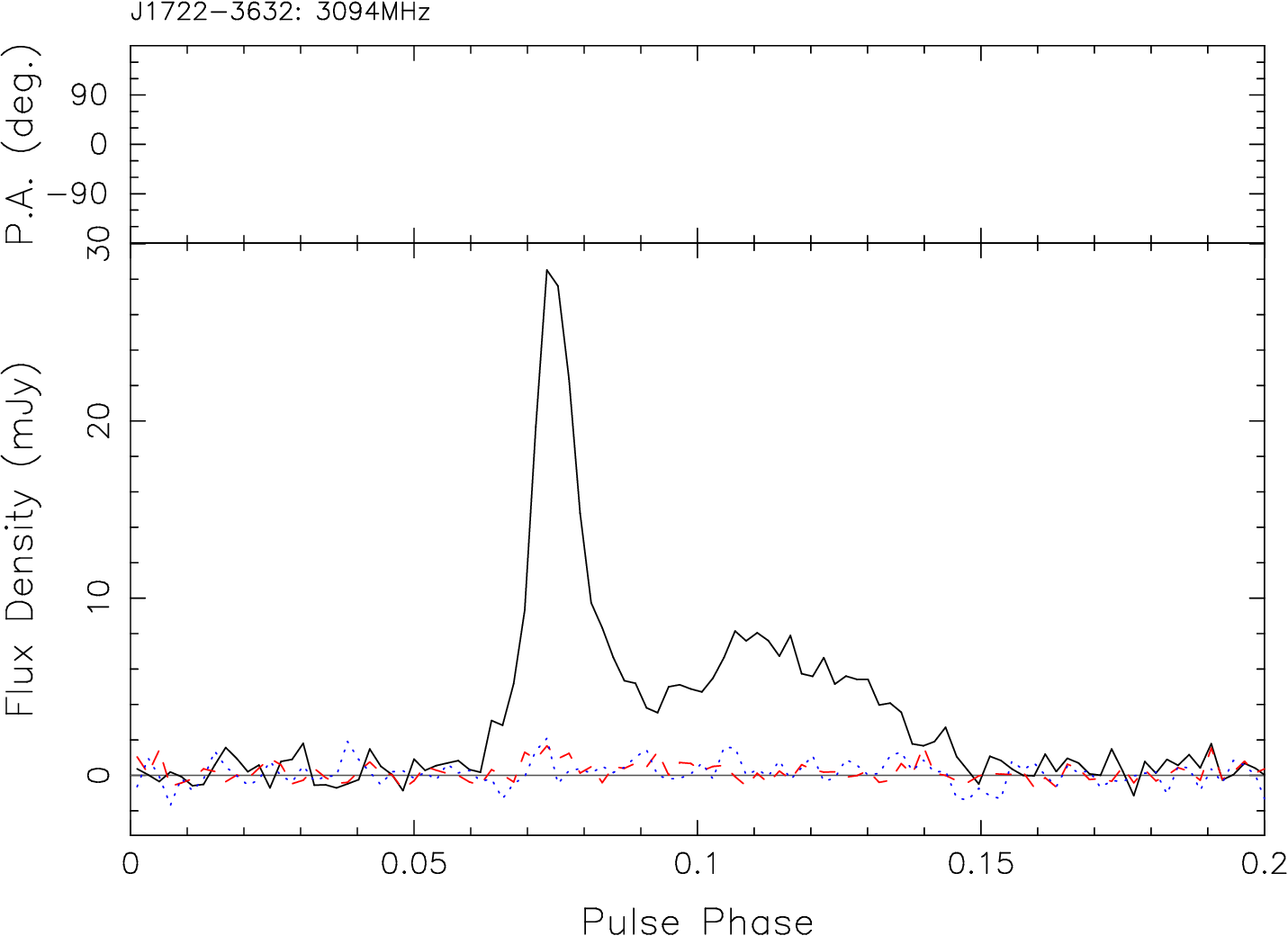}}& 
\resizebox{0.4\hsize}{!}{\includegraphics[angle=-90]{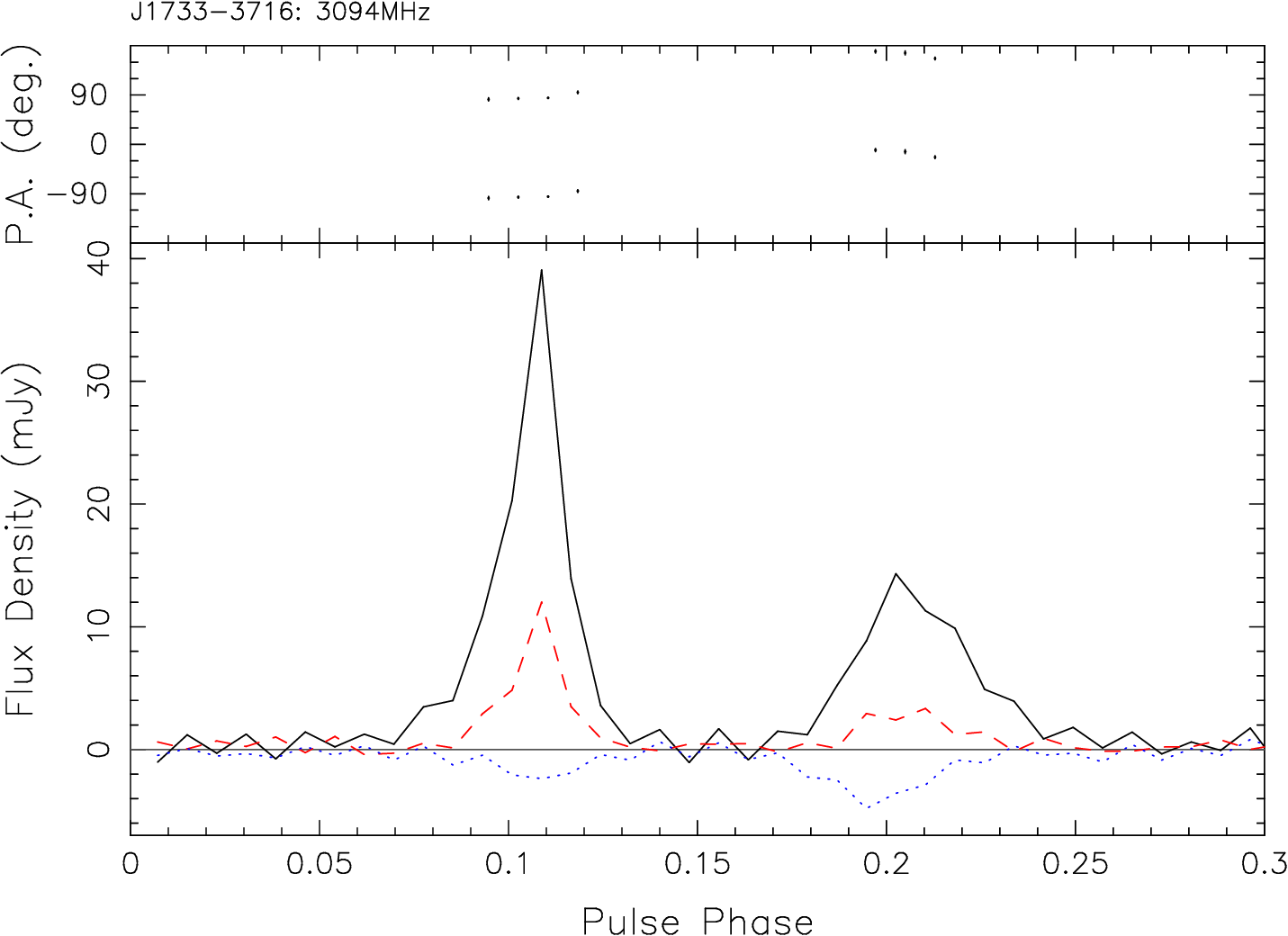}}\\
\end{tabular}
\end{center}
\caption{- continued.}
\end{figure*}
\addtocounter{figure}{-1}
\begin{figure*}
\begin{center}
\begin{tabular}{cc}
\resizebox{0.4\hsize}{!}{\includegraphics[angle=-90]{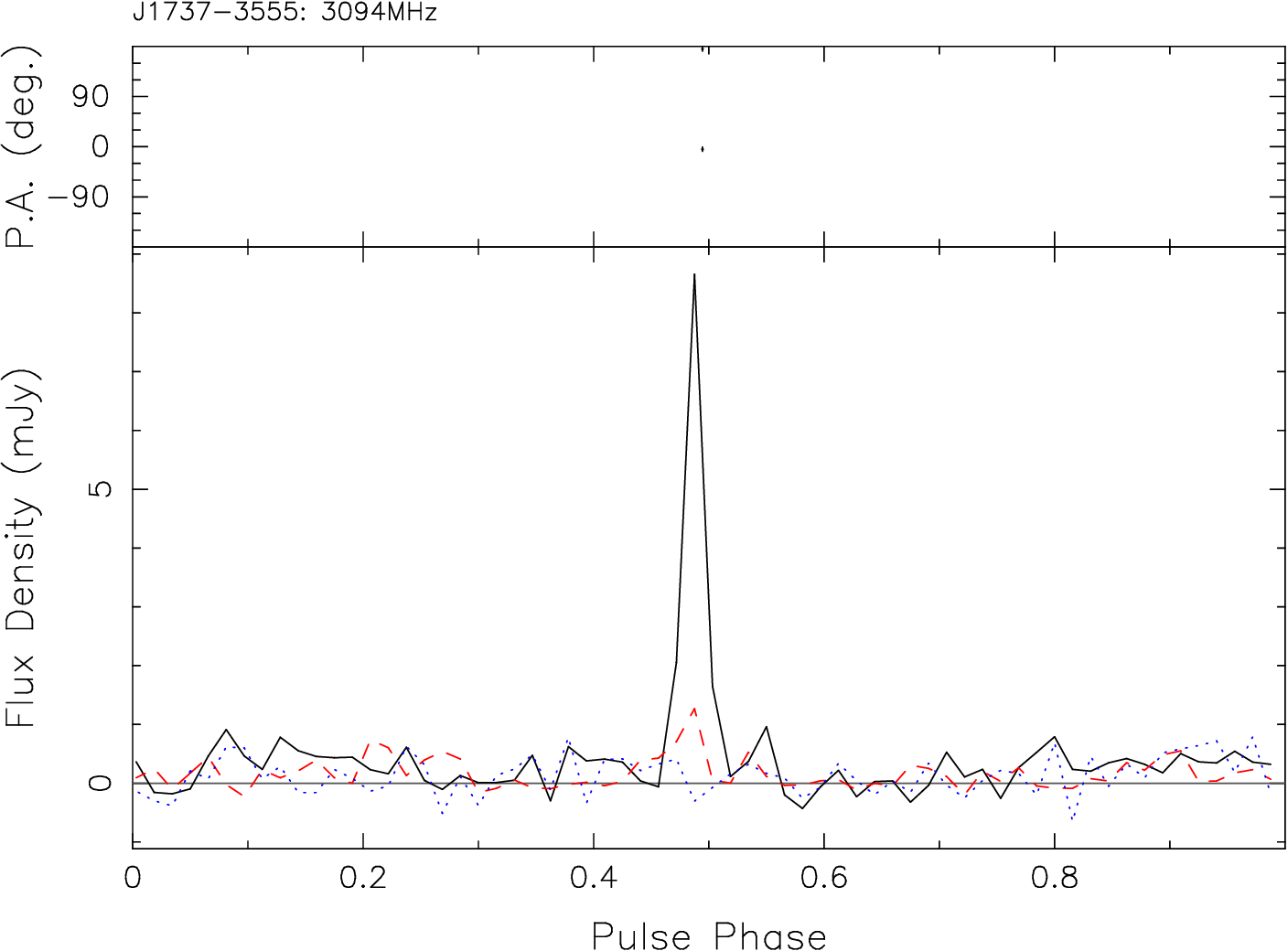}}& 
\resizebox{0.4\hsize}{!}{\includegraphics[angle=-90]{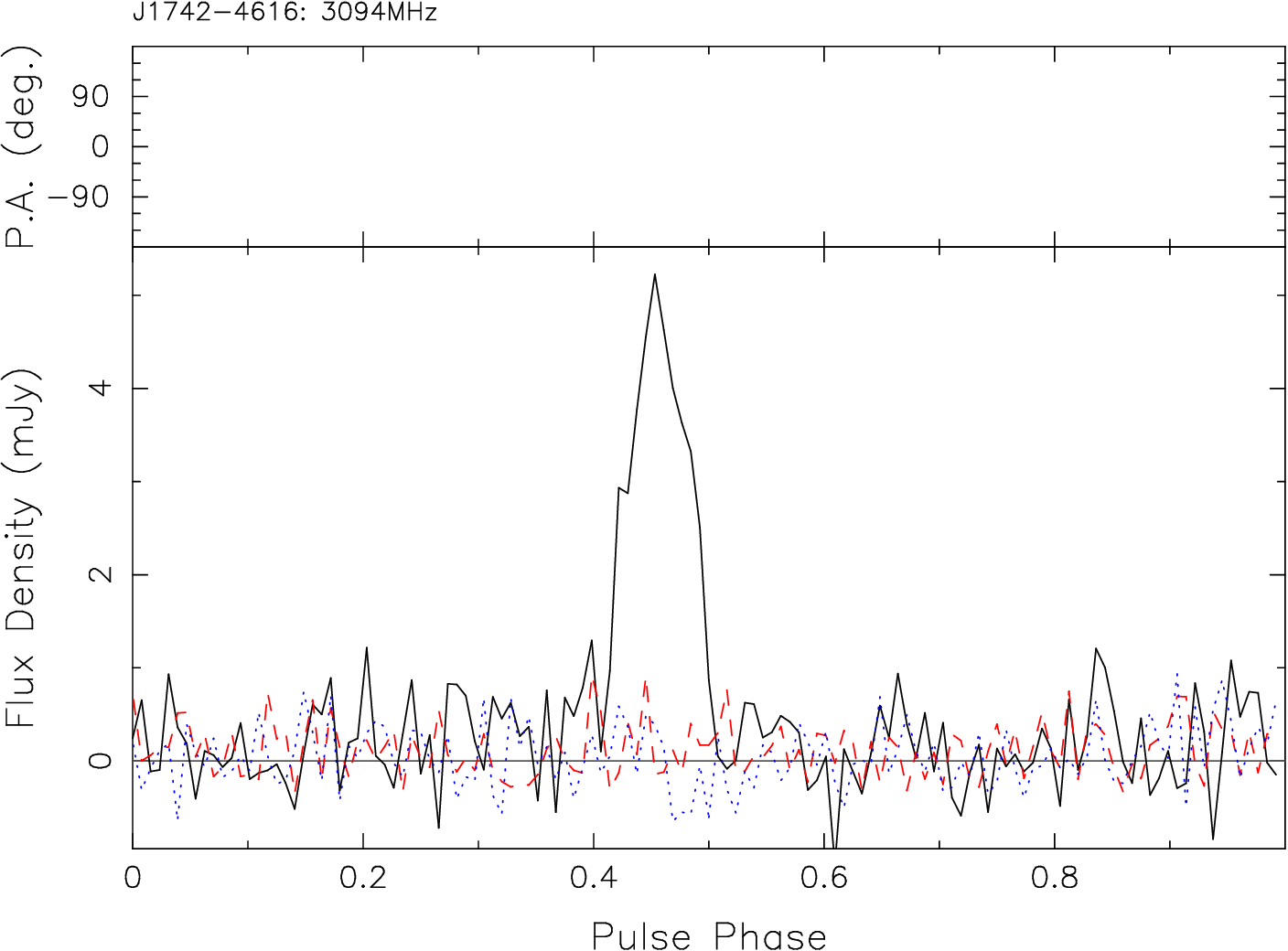}}\\
\resizebox{0.4\hsize}{!}{\includegraphics[angle=-90]{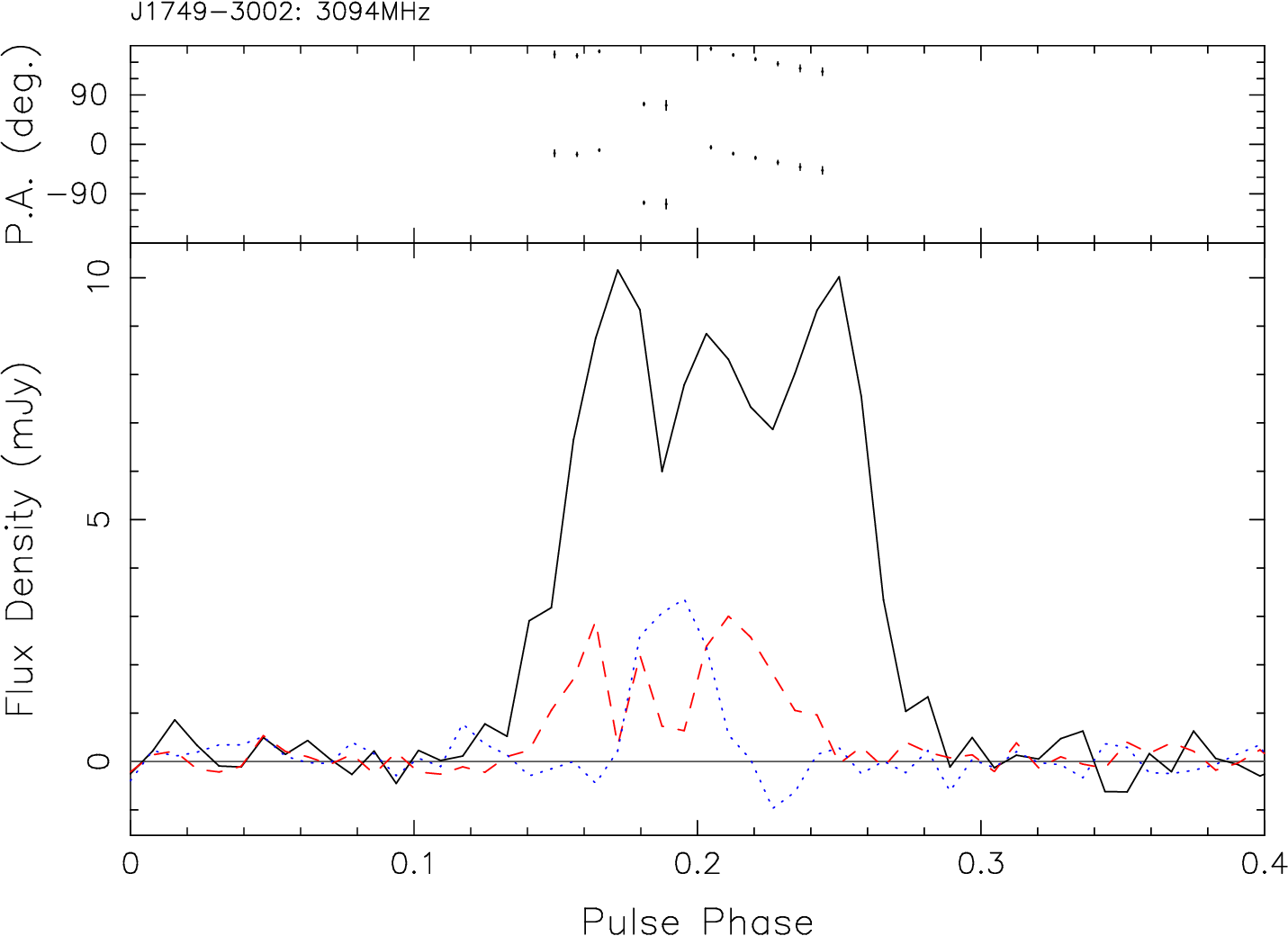}}& 
\resizebox{0.4\hsize}{!}{\includegraphics[angle=-90]{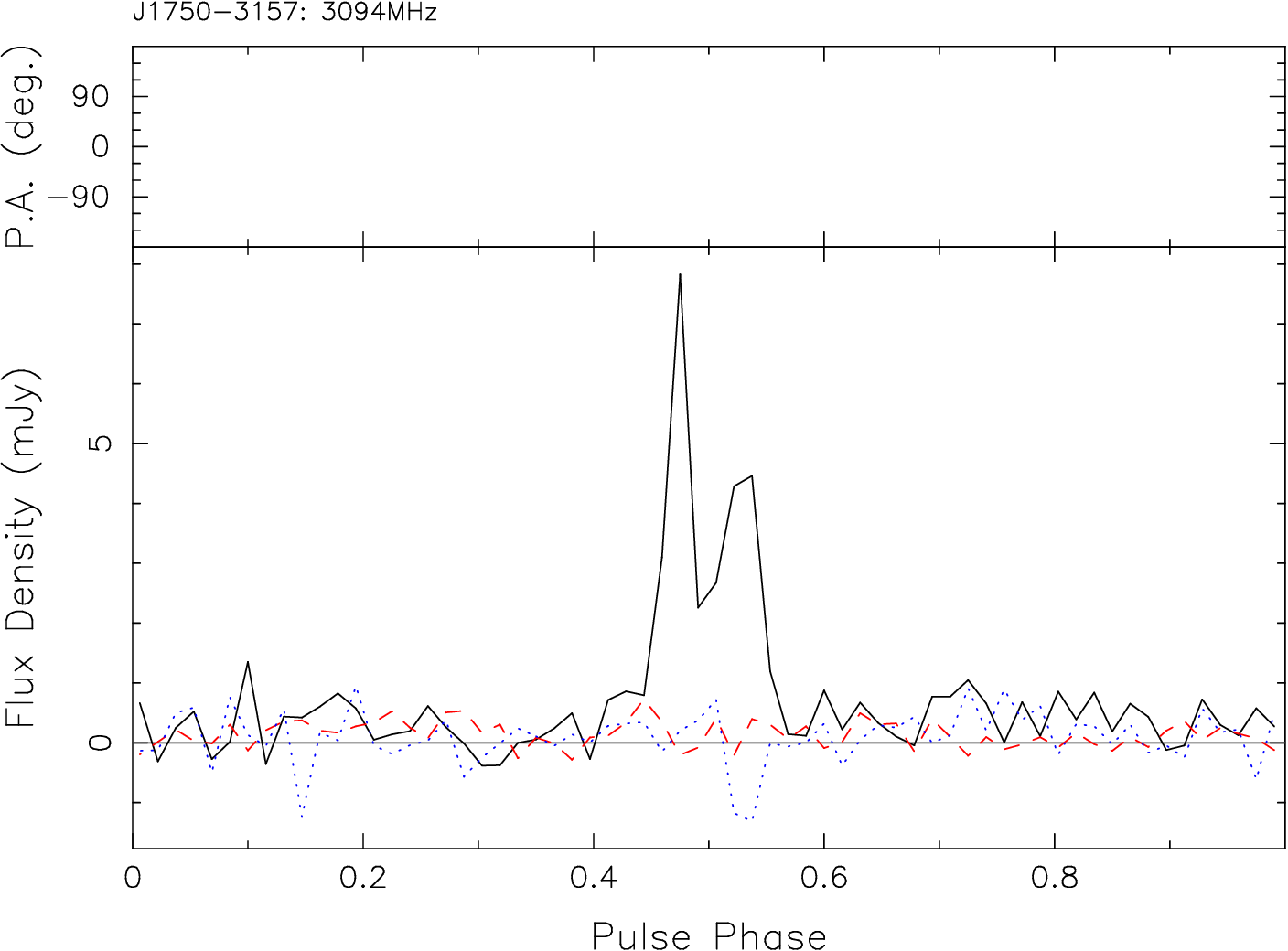}}\\
\resizebox{0.4\hsize}{!}{\includegraphics[angle=-90]{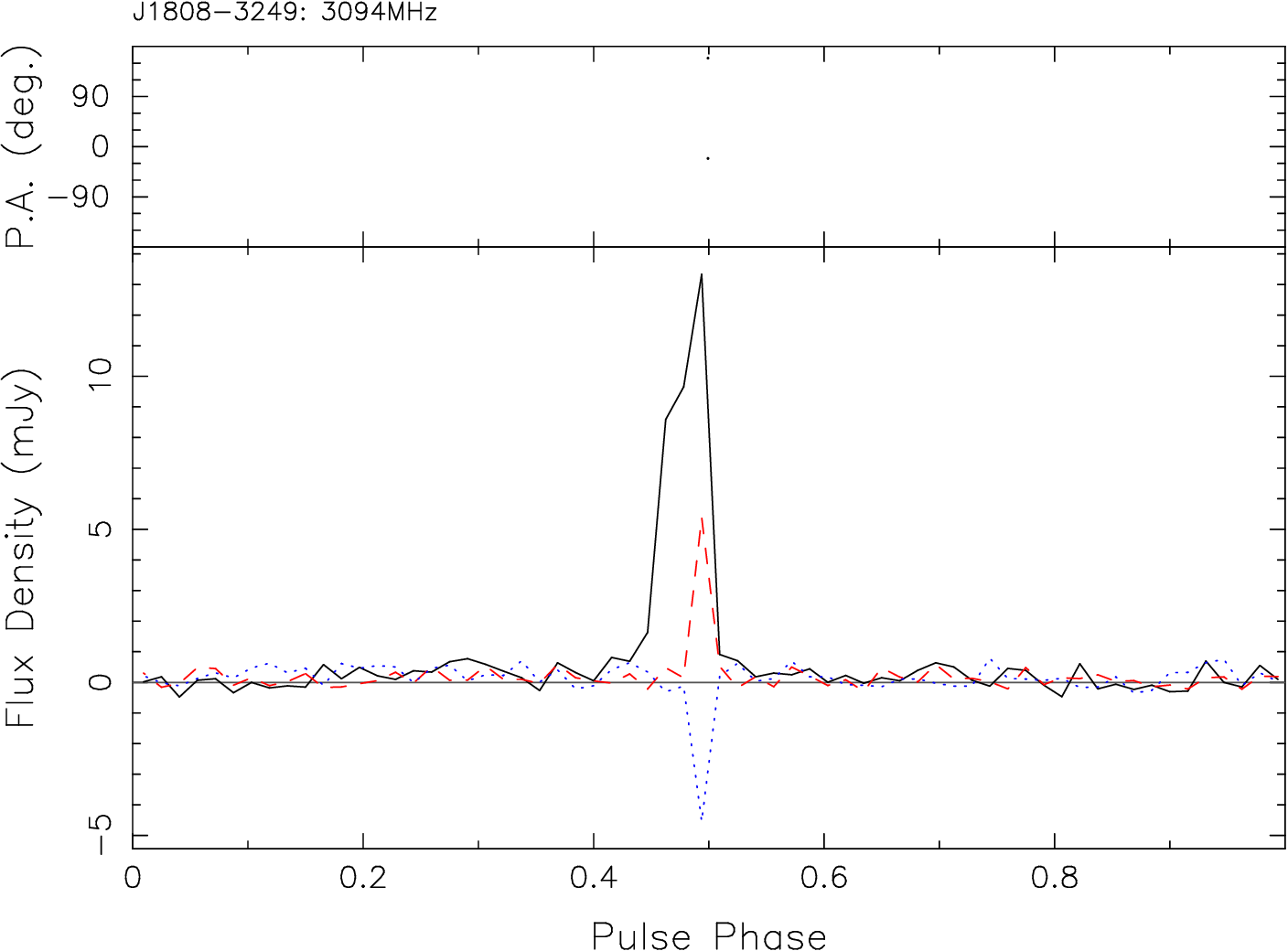}}& 
\resizebox{0.4\hsize}{!}{\includegraphics[angle=-90]{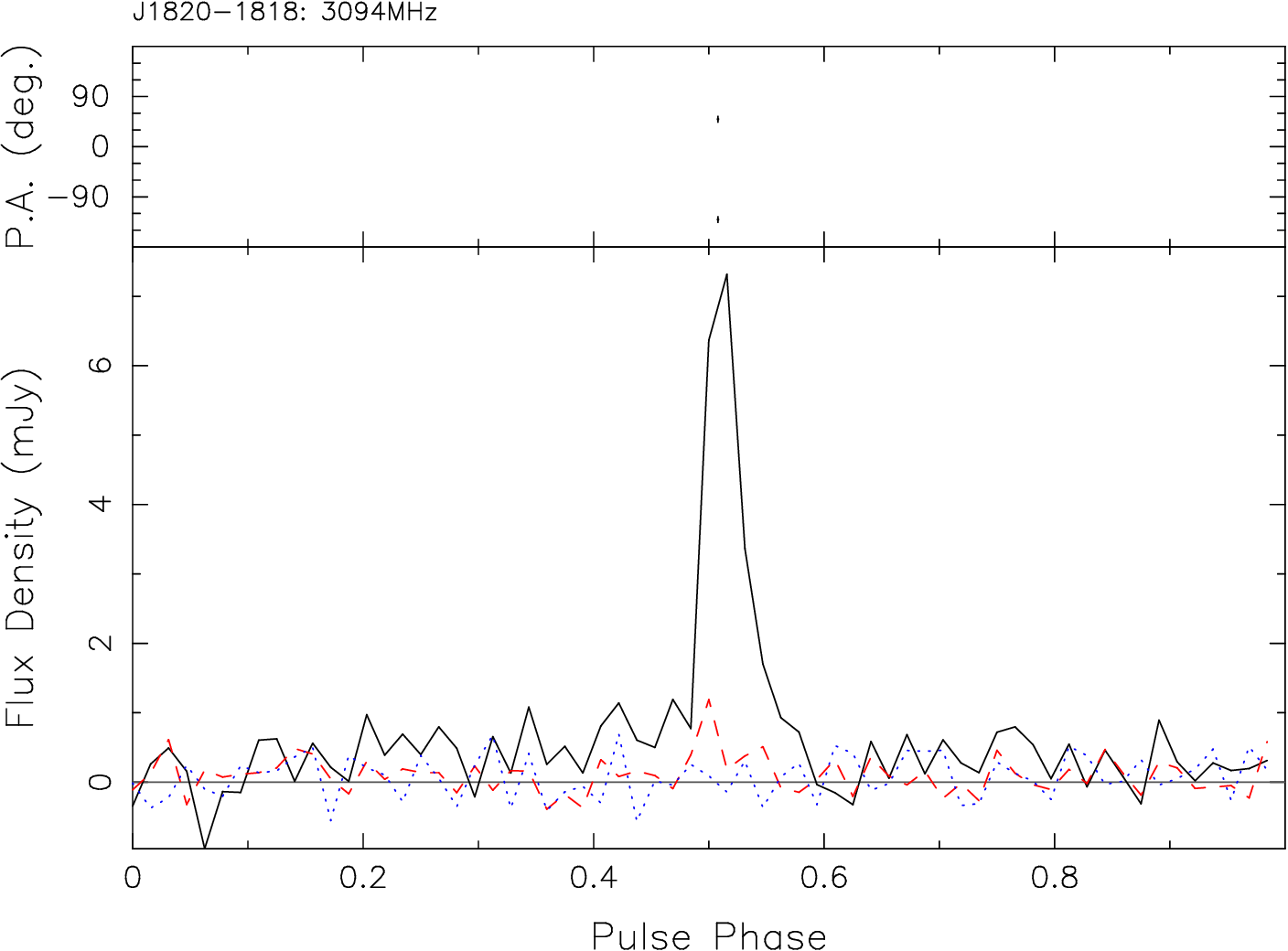}}\\
\resizebox{0.4\hsize}{!}{\includegraphics[angle=-90]{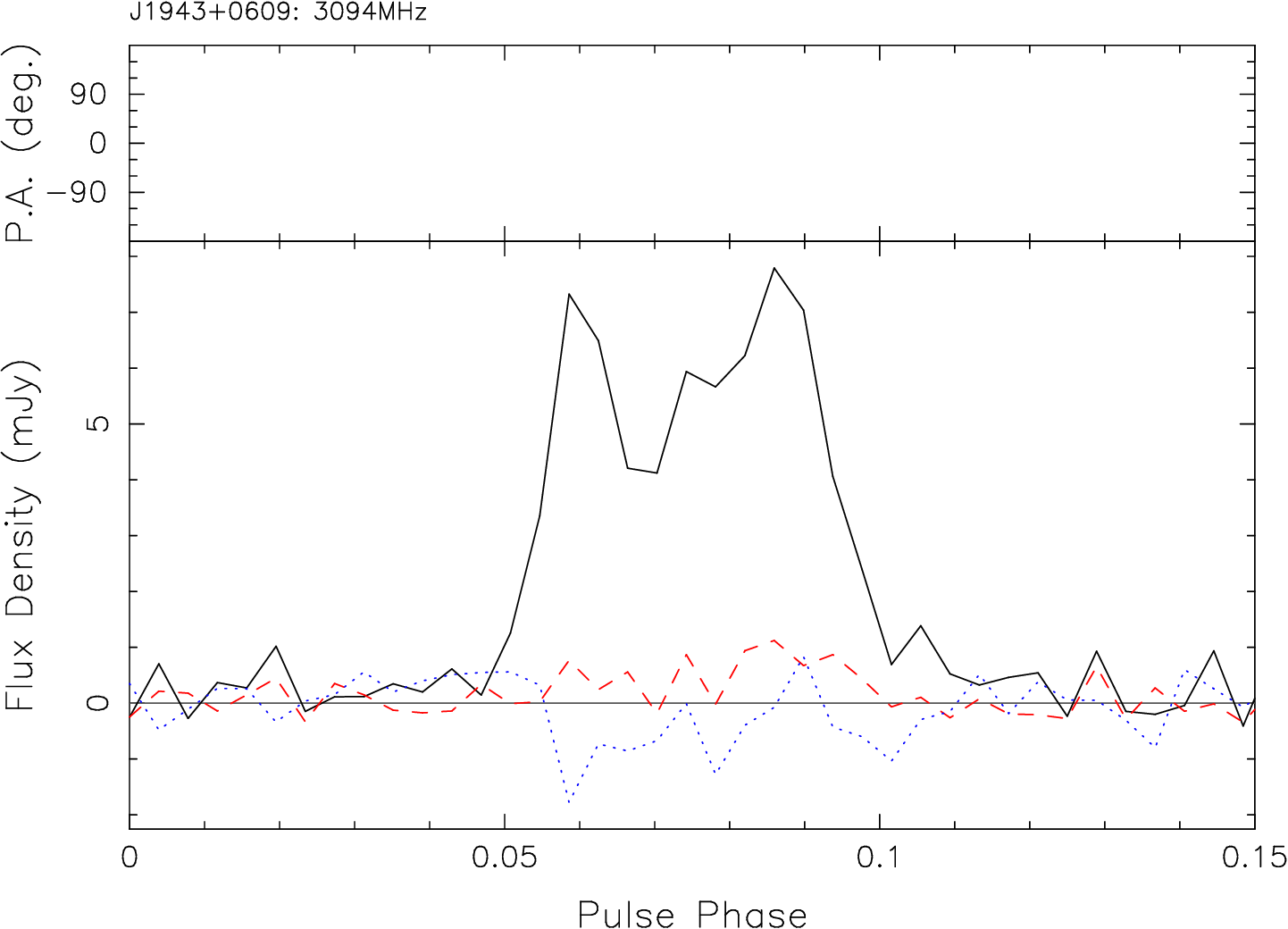}}& 
\resizebox{0.4\hsize}{!}{\includegraphics[angle=-90]{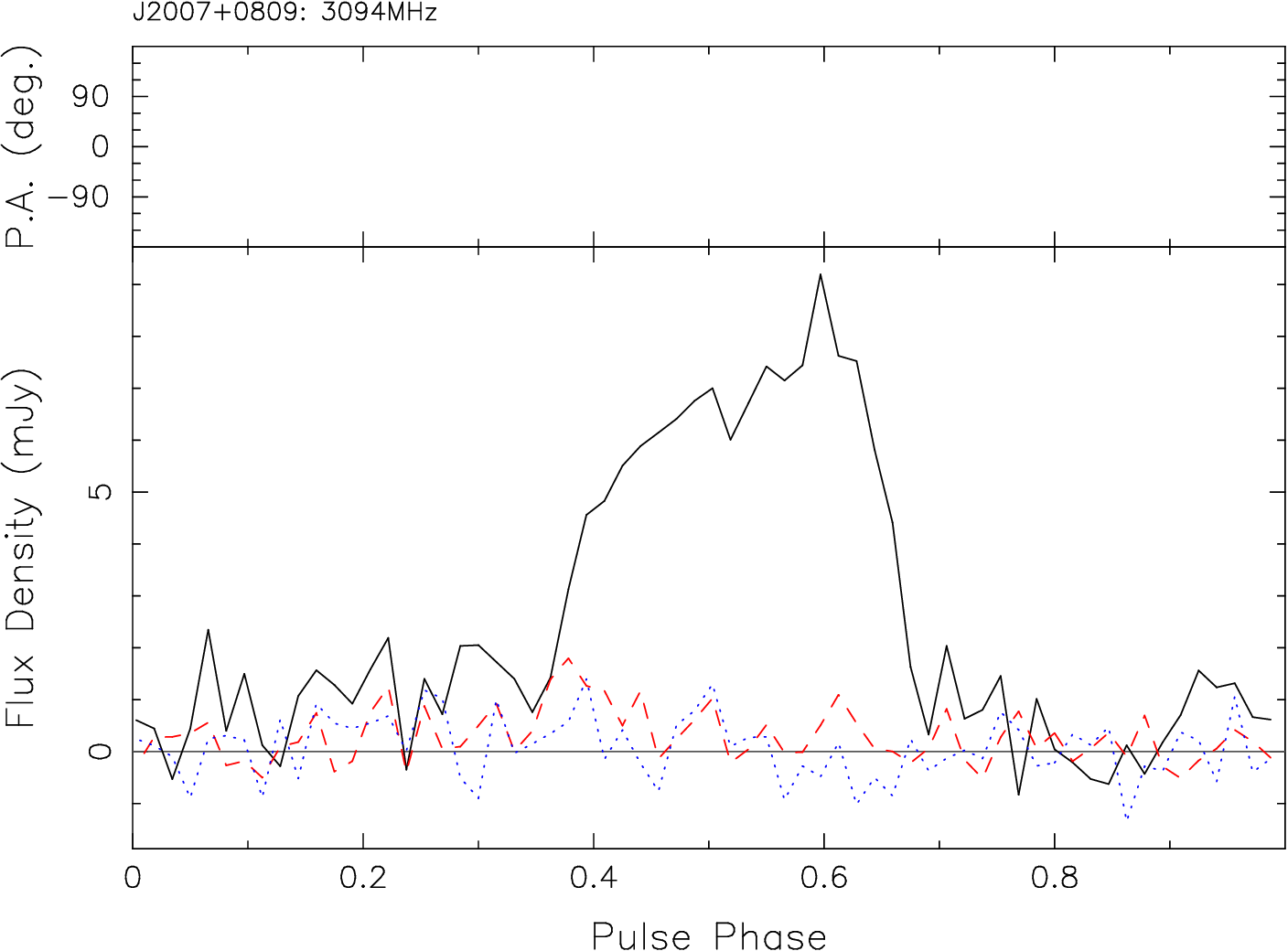}}\\
\end{tabular}\\
\end{center}
\caption{- continued.}
\end{figure*}

\section{Polarimetric Profiles}

The pulsars we observed at 3.1 GHz are listed in Table 1 along with
their periods and continuum flux densities at 1.4 GHz \citep{hfs+04}
and 3.1 GHz. They are generally weak at 3.1 GHz, the brightest being
J1721$-$3532 with a flux density of 6.4 mJy. The width and average
degree of linear polarisation as a fraction of the continuum flux
density is also given in the table. However, due to the different
dependence on frequency, important changes in the polarisation of
individual components may not be paid the attention they merit when
using these average values. We present the polarimetric profiles at
3.1 GHz in Figure~1. In the following subsections, we describe the
profiles in detail, starting with the highly polarised sources and
ending with the sources with low polarisation.

\subsection{Highly polarised sources}

{\bf J1048$-$5832.} This young pulsar was observed at 1.4 GHz by QMLG,
and exhibits a high degree of linear polarisation and almost no
circular. The 3.1 GHz profile has, somewhat surprisingly, even higher
linear polarisation than at 1.4 GHz, contrary to the usual
de-polarisation observed with increasing frequency. There is also
significant circular polarisation at 3.1 GHz. It would be useful to
follow up this unexpected behaviour with observations at higher
frequencies.

{\bf J1302$-$6350.} This pulsar is well known to have a highly
polarised, wide, double profile \citep{mj95}. The 3.1 GHz profile
falls nicely between the 2.2 and 4.8 GHz profiles displayed in
\citet{wjm04}. Although both components are virtually completely
linearly polarised at all frequencies, there is a notable increase in
the linear polarisation of the strongest of the two components above
1.5 GHz. The PA profiles at all frequencies are similar, with a
negative slope in each component.

{\bf J1539-5626.} The profile at 3.1 GHz is complex and consists of at
least two components. The trailing component is very highly linearly
polarised with a flat PA. There is a swing in the average circular
polarisation from right handed in the middle, brightest component, to
left handed in the trailing component and the pulse phase of the change
in sense coincides with a minimum in linear polarisation. The poor
temporal resolution of the 1.4 GHz profile (QMLG) precludes an
accurate comparison, but it appears the trailing component is less
polarised at the lower frequency.

{\bf J1614$-$5048.} At 3.1 GHz the profile consists of two components,
both of which are highly linearly polarised. Relatively high circular
polarisation is seen in the trailing component. The PA profile has a
negative slope across the two bright components. The profile at 1.4
GHz shows only one component (QMLG), which we associate with the
trailing component of the 3.1 GHz profile. Intermediate and higher
frequency observations of this pulsar are warranted to study the
frequency evolution.

{\bf J1637$-$4553.} The profile of this pulsar, which consists of a
single component, exhibits a high degree of linear polarisation at 3.1
GHz, with a flat PA across the pulse. The swing in PA at the trailing
edge is common between this profile and unpublished data at 1.4 GHz.
The profile also appears narrower in linear polarisation than in total
power: the fractional linear polarisation is significantly less at the
edges of the profile than in the middle. A small fraction of
right-hand circular polarisation is seen.

{\bf J1709$-$4429.} High linear polarisation and moderate,
right-handed circular polarisation are observed in the mean profile of
this young pulsar at 3.1 GHz. The wide single-component pulse is
identical to the 1.4 GHz profile in QMLG, with a smooth PA profile of
positive slope.

\subsection{Moderately polarised sources}

{\bf J1038$-$5831.} The linear polarisation is high enough to permit a
determination of the PA across most of the pulse. The kinkiness in the
PA profile of QMLG at 1.4 GHz can be seen in the 3.1 GHz profile,
where it resembles an orthogonal jump. More specifically, just before
the peak of the brightest component, the linear polarisation has a
sharp dip which occurs simultaneously with a jump of the PA. A
comparison of the linear polarisation profiles at 3.1 GHz and 1.4 GHz
reveals some de-polarisation at the leading edge of the pulse.

{\bf J1319$-$6056.} The 3.1 GHz profile suggests moderately high
degrees of linear and circular polarisation, similar to the 1.4 GHz
profile of QMLG.

{\bf J1413$-$6307.} QMLG remark that this pulsar shows no linear or
circular polarisation at 1.4 GHz. However, at 3.1 GHz the profile
exhibits moderate linear polarisation towards the trailing edge of the
pulse. The mean circular polarisation shows a swing of sense from
left- to right-handed. The PA follows a smooth curve across the pulse.

{\bf J1522$-$5829.} The leading component of the 3.1 GHz profile is
substantially more polarised than its counterpart at 1.4 GHz in
QMLG. At 3.1 GHz the PA profile consists of two flat segments, $\sim
70\degr$ offset from each other, whereas, at 1.4 GHz, the PA profile
shows a constant negative slope in both segments. Such differences
imply a degree of independence of the PA from the geometry of the
pulsar and point towards propagation effects in the pulsar
magnetosphere.

{\bf J1535$-$4114.} The moderately high linear polarisation in this
pulse profile at 3.1 GHz is characterised by a PA curve with a steep
positive gradient. A small amount of left-hand circular polarisation
can also be discerned.

{\bf J1630$-$4733.} At 1.4 GHz, the profile of this pulsar is severely
scattered. At 3.1 GHz, the high average right-hand circular
polarisation resembles the total power profile. The linear
polarisation consists of two components, with a small swing at the
pulse phase of the local minimum in linear polarisation.

{\bf J1633$-$5015.} At 1.4 GHz, the single-component profile of this
pulsar has both high linear and circular polarisation (QMLG). At 3.1
GHz, this component has less linear polarisation. The profile also has
a leading component, not seen at 1.4 GHz, which is unpolarised. An
orthogonal PA jump can be inferred at the leading edge of the profile,
whereas the PA profile is flat over the middle component.

{\bf J1646$-$4346.} This weak pulsar has a profile which consists of
multiple components at 3.1 GHz. The degree of linear polarisation is
highest at the leading and trailing edges of the pulse, although it
remains moderately high under the pulse peak. The PA profile is linear
with a negative slope. There is a small amount of right-hand circular
polarisation. This pulsar was observed at 1.4 GHz by
\citet{cmk01}. Although weak, it shows similar characteristics
to the higher frequency profile with a somewhat higher degree of
linear polarisation.

{\bf J1653$-$3838.} The profile of this pulsar at 3.1 GHz shows a
moderate amount of linear polarisation in the trailing component,
characterised by a slightly rising PA. The leading component is not
linearly polarised and the circular polarisation is not significant
across the entire pulse. At 1.4 GHz the profile is similar, although
there is a high degree of circular polarisation in the trailing
component.

{\bf J1707$-$4053.} The 3.1 GHz profile is characterised by moderate
linear polarisation and circular polarisation that swings from right
to left handedness. The gradient of the PA profile changes sign at
approximately the pulse phase of the profile peak. At 1.4 GHz, the
pulse profile is very wide and the PA is flat (QMLG), both most likely
caused by interstellar scattering.

{\bf J1721$-$3532.} The total power profile of this pulsar at 3.1 GHz
consists of two overlapping components, both with moderately high
linear polarisation. Both components can be identified in the
scattered 1.4 GHz profile (QMLG), but the trailing one is less
linearly polarised at that frequency. The PA profile consists of two,
approximately linear segments with a negative slope, each apparently
associated to a profile component. There is moderate right-hand
circular polarisation across the pulse.

{\bf J1733$-$3716.} Each of the two components of this profile at 3.1
GHz is characterised by a moderate degree of linear polarisation and
some right-hand circular polarisation. The PA has a slightly positive
slope in the leading component and a slightly negative slope in the
trailing component. At 1.4 GHz, the profile looks similar.

{\bf J1749$-$3002.} A comparison of the total power profile at 3.1 GHz
to the 1.4 GHz profile in QMLG, shows that the outriders have flatter
spectra than the middle component. At 3.1 GHz, there are two
orthogonal PA transitions which coincide with the minima in linear
polarisation. The circular polarisation shows a left-handed maximum
between the leading and middle components.

{\bf J1808$-$3249.} The profile of this weak pulsar at 3.1 GHz shows
equal amounts of linear and circular polarisation in the trailing
component. 

\subsection{Sources with low polarisation}

{\bf J1012$-$5857.} At 1.4 GHz, the profile of this pulsar shows no
measurable linear and circular polarisation (QMLG). However, at 3.1
GHz, the measured linear polarisation is sufficient to provide a PA
profile across the central part of the profile. The PA exhibits a
discontinuity just prior to the peak of the total power profile, where
the linear polarisation profile also shows a local minimum.

{\bf J1110$-$5637.} At 3.1 GHz there is measurable linear polarisation
in the leading and middle parts of the profile and circular
polarisation in the trailing component. The PA has a positive slope in
both of these components, with a quasi-orthogonal discontinuity. The
1.4 GHz profile of QMLG has poor resolution and signal-to-noise
but better quality (unpublished) data shows the 1.4 and 3.1 GHz profiles
to be similar.

{\bf J1114$-$6100.} The profile of this pulsar at 3.1 GHz resembles
the profile at 1.4 GHz. It consists of a single, total power component
with no measurable linear and circular polarisation.

{\bf J1126$-$6054.} The total power profile of this pulsar at 3.1 GHz
has more pronounced ``outriders'' than the 1.5 GHz profile in
\citet{jlm+92}. There is a small degree of linear and circular
polarisation across the profile. On average, the circular polarisation
is right-handed in the leading part of the profile and left-handed in
the trailing part of the profile.

{\bf J1133$-$6250.} This pulsar shows no detectable polarisation at
3.1 GHz and is similar to the 1.4 GHz profile in QMLG, except for the
fact that leading component is brighter with respect to the
trailing component at 3.1 GHz.

{\bf J1306$-$6617.} There is an orthogonal jump in the PA profile of
this pulsar at 3.1 GHz, at the trailing edge of the pulse. Both the
total power and the polarisation at 3.1 GHz are similar to the 1.4
GHz profile in QMLG, with minor differences mainly due to different
temporal resolution.

{\bf J1327$-$6301.} The linear polarisation of the pulsar at 3.1 GHz
shows a number of local minima. At the pulse phase of the second
minimum, there is an orthogonal PA jump. Both linear and circular
polarisation are otherwise low across the pulse. At 1.4 GHz, the total
power profile looks the same, however, there is significant circular
polarisation that swings from left to right handedness.

{\bf J1338$-$6204.} This profile consists of a middle component
flanked by two outriders. Both the linear and circular polarisation
profiles appear symmetrical about the peak of the middle component. A
comparison with the 1.4 GHz profile of QMLG demonstrates flatter
spectra in the outriders with respect to the middle component. It also
reveals an increase in linear polarisation of the trailing
component. Also, the PA appears to be ascending in the leading
component and descending in the trailing component.

{\bf J1410$-$7404.} The profile of this pulsar is very narrow at 3.1
GHz and only slightly linearly polarised. No significant circular
polarisation was observed.

{\bf J1512$-$5759.} The simple total-power pulse is characterised by
an absence of
linear polarisation at 3.1 GHz. There is a hint that the average
circular polarisation may be changing sense shortly after the middle
of the pulse.

{\bf J1611$-$5209.} At 3.1 GHz, there is an inter-pulse in the
profile, exactly $180\degr$ away from the strongest peak. There is little
linear and circular polarisation detected in the main pulse and none
in the inter-pulse. At 1.4 GHz, the linear polarisation of the main
pulse is higher and the PA shows a significant swing.

{\bf J1640$-$4715.} The double component profile of this pulsar at 3.1
GHz shows some linear polarisation in the leading, weak component and
no linear or circular polarisation in the strong component. The two
measured values of PA in the leading component indicate a positive
slope.

{\bf J1655$-$3048.} Despite being weak at 3.1 GHz, significant left-
and right-hand average circular polarisation is observed in this
profile. The pulse phase of the change in sense approximately
coincides with the peak of the profile. Some linear polarisation is
observed towards the trailing edge of the profile.

{\bf J1701$-$4533.} The profile of this pulsar at 3.1 GHz resembles
the 1.4 GHz profile in QMLG, in that it exhibits small detectable
linear and circular polarisation. The PA swing has a negative slope.

{\bf J1722$-$3632.} At 3.1 GHz, this double component profile shows no
significant linear polarisation or circular polarisation.

The following pulsars are all weak at 3.1 GHz and their profiles
exhibit very little or no detectable polarisation: J1352$-$6803,
J1517$-$4356, J1534$-$5405, J1615$-$5537, J1633$-$4453, J1712$-$2715,
J1719$-$4006, J1737$-$3555, J1742$-$4616, J1750$-$3157, J1820$-$1818,
J1943+0609, J2007+0809.

%
\section{Discussion}
\begin{figure}
\begin{center}
\resizebox{\hsize}{!}{\includegraphics{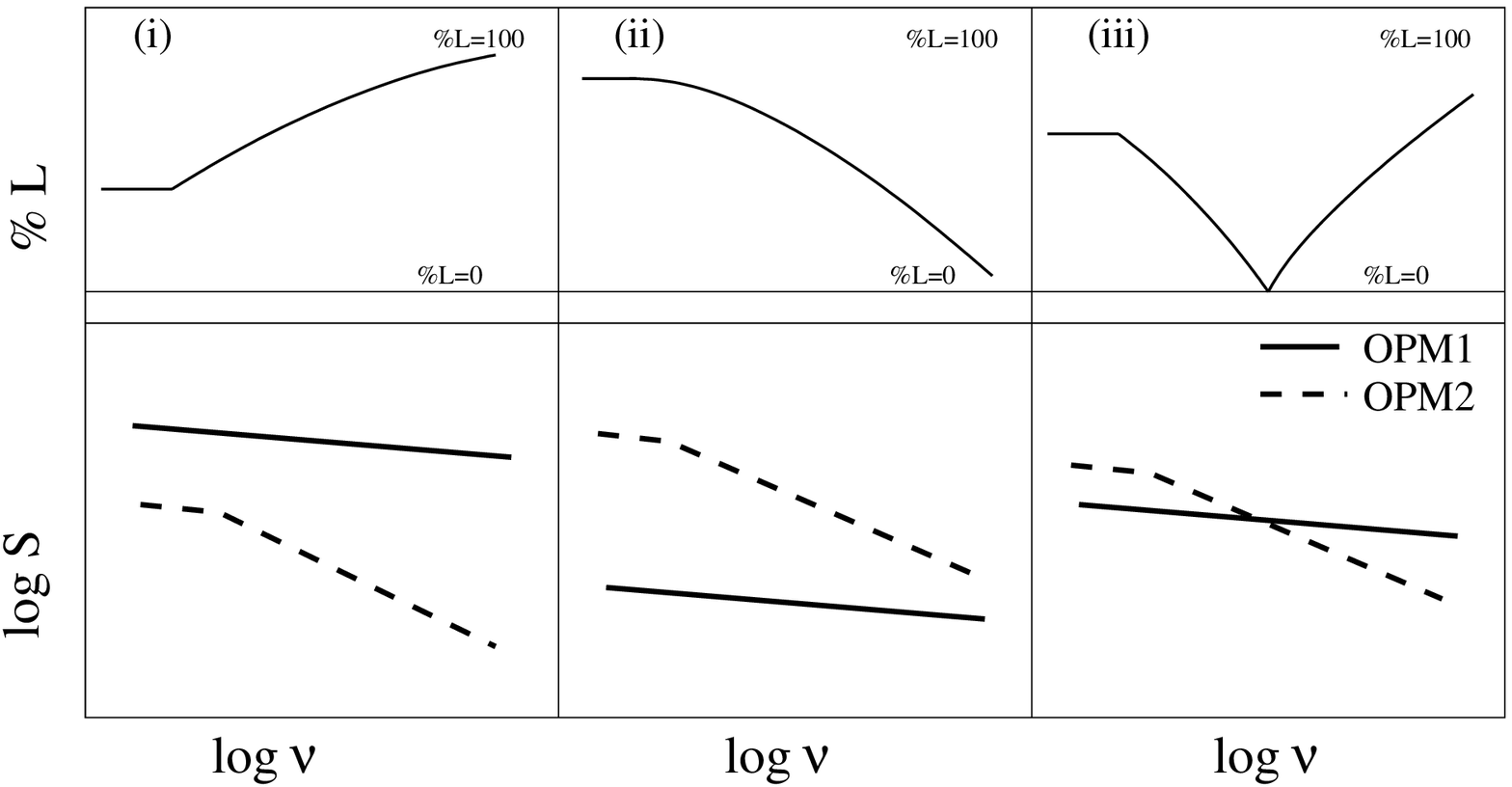}}\\
\end{center}
\caption{Three possibilities for the spectral behaviour of OPM in
  individual components within a frequency window, where $S$ is the
  flux density and $\nu$ the frequency. The top panels represent the
  degree of linear polarisation resulting from the superposition of
  the OPM. }
\end{figure}

The 48 profiles shown in Figure~1 demonstrate the diversity of
polarisation properties seen in pulsars. However, systematic features
are also present in the data.  A well--known example is the fact that
abrupt changes in the PA - often approximately orthogonal jumps -
occur at the pulse phase of local minima in the linear
polarisation. This suggests that one cause of the general
de-polarisation with increasing frequency is the superposition of
linearly polarised rays with different planes of polarisation.  Also,
our observations re-enforce the fact that different components of
pulse profiles often exhibit different spectral behaviour. This leads
to changes in the overall shape of profiles with frequency beyond that
of simple radius--to--frequency mapping expectations, as detailed in
\citet{kra94}.

We use our results in the context of the model for polarised emission
proposed by \citet{ms98}. In that model, the emission consists of two
completely polarised modes which are orthogonal to each other. If
circular polarisation is ignored then the superposition of these modes
at any instant yields the total power (the sum of the intensity of the
modes) and the linear polarisation (the difference of the mode
intensities). Clearly then, if the mode intensities are roughly equal
the linear polarisation is very low; if they are very disparate the
linear polarisation will be close to 100 per cent. The question then
arises as to how the relative intensities of the modes varies as a
function of frequency and/or component type \citep{kkj+02}.

We propose a component classification scheme based on three possible
scenarios for the spectral index behaviour of the modes as shown
schematically in Figure~2. Note that the sketches are conceptually
similar; one mode has a relatively flat spectrum and the other a
relatively steep spectrum, that flattens off at low frequencies. In
more detail, the scenarios are as follows.
\begin{enumerate}
\item The intensity of the flat spectrum mode is always greater than
  the steep spectrum mode.  In this case, the difference between the
  strengths of the OPM will increase with increasing frequency, which
  results in higher degrees of linear polarisation at higher
  frequencies.  The total power spectral index is dominated by the
  mode with the flatter index. Examples from the current work are PSRs
  J1048$-$5832 and J1539$-$5626 (high polarisation), J1413$-$6307,
  J1522$-$5829 and J1721$-$3532 (moderate polarisation) and
  J1012$-$5857 and J1338$-$6204 (low polarisation), where an increase
  of the degree of linear polarisation is seen in individual
  components at 3.1~GHz compared to 1.4~GHz. PSRs J1302$-$6350,
  J1614$-$5048, J1637$-$4533 and J1709$-$4429 are examples of pulsars
  in this category which have virtually 100 per cent linear
  polarisation over a large frequency range. Support for this picture
  also comes from the fact that many pulsars in this category have
  rather flat spectral indices, especially those with
  single--component profiles. The average spectral index of pulsars is
  $-$1.8 \citep{mkkw00a}; in the current sample, PSRs J1048$-$5832,
  J1302$-$6350, J1413$-$6307, J1709$-$4429 and J1721$-$3532 all have
  spectral indices flatter than $-$1.0, two pulsars, PSRs J1614--5048
  and J1637--4553 have spectral indices of $-$1.3 and $-$1.6 and the
  remaining have complex profiles making it hard to determine
  individual component spectra.
\item The intensity of the steep spectrum mode is initially greater
  than the flat spectrum mode and the intersection of the two spectra
  occurs at a high frequency. The consequence of this is high linear
  polarisation at low frequencies, decreasing towards higher
  frequencies as a power-law. Decreasing polarisation with frequency
  is commonly observed and \citet{xsg+95} present evidence for
  power-law behaviour of fractional polarisation. Under this scenario
  we predict that the polarisation will eventually increase again at
  some high frequency.
\item The spectra cross at a frequency near $\sim$1 GHz, within the
  window where the bulk of current observations lie.  Then, when
  observing from low to high frequencies, the linear polarisation
  should decrease to zero at the frequency of the intersection and
  increase again beyond that frequency with an associated orthogonal
  PA jump. The total power spectrum flattens after the intersection
  point. There exist a number of pulsars with components that show
  orthogonal PA jumps at around 1~GHz and as such are candidates for
  this scenario. We intend to investigate their polarisation behaviour
  over a wide frequency range.
\end{enumerate}

A classification scheme for the polarisation of pulsar profiles was
proposed by \citet{hkk98}, wherein they recognised examples that
constitute both type (i) and type (ii) cases. However, their scheme
was based on the global profile properties rather than individual
components and was not based on the properties of orthogonally
polarised modes. Their observations of PSR B0355+54 at frequencies up
to 32~GHz strongly support our picture. In that pulsar, the first
component remains highly polarised at all frequencies (i.e. it is of
type i), whereas the second component has decreasing polarisation (it
is of type ii). The first component has a flatter spectral index than
the second component and begins to dominate the profile above 5~GHz
exactly as expected in our simple model. Taking the profile as a whole
\citep[e.g.][]{kxj+96}, the spectrum {\it apparently} shows a break
around 5~GHz but this is simply due to the difference in spectral
index between the two components and shows the importance of treating
components individually.

Counter-examples to our model may come from the high frequency
observations of \citet{xkj+96}. They present some evidence that the
polarisation fraction appears to have a spectral break in some pulsars
(e.g PSRs B0329+54 and B1133+16).  A direct comparison with our ideas
is however difficult as our model is based on individual components
whereas their study treated the pulse profile as a whole. In any case,
it has been mooted that other factors, such as a loss of coherence
\citep{xkjw94}, are behind the polarisation effects that are seen at
frequencies above 10~GHz, which our model does not account for.

The approach we use is based on the linear polarisation and total
power spectra resulting from superposed OPM. In this model, the degree
of linear polarisation and the total power are therefore tied at a
given frequency. The fact that the highly linearly polarised pulsars
described in Section 3.1 have flatter spectra than the average (the
spectral indices range from $-0.27$ to $-1.56$, compared to the
average $-1.6$) is a good indication of this tie. OPM intensities are
frequency dependent, which is well documented from previous studies
\citep{kkj+02}, and the theoretical interpretations largely involve
propagation effects in the magnetosphere. Our challenge to this
picture is if and where there is a well defined frequency at which the
OPM spectra cross. We expect different physical conditions determining
the OPM spectra for components in the different scenaria of our
model. Those differences may be due to the location of the component
in the emission region or other parameters, such as age or spin-down
energy.

The geometrical elegance of the rotating vector model and its
predictions about PA profiles \citep{rc69a} provide a basis for
understanding the profiles observed. A consequence of the PA being
determined purely by geometry is that the PA profiles should be
identical at all observing frequencies (apart from relativistic
effects, see \citealt{ml04}).  A known deviation from the pure
geometrical interpretation of the PA comes from the frequency
dependence of OPM, with the possibility of $90\degr$ offsets in PA
between parts of the profile at different frequencies. Also, the PA
tends to deviate from the simple model at pulse phases where
individual components overlap.  In the data presented here, we
identify two pulsars with notable differences in the PA profiles that
go beyond simple orthogonal (or even non-orthogonal) deviations. PSRs
J1522$-$5829 and J1707$-$4053 show complicated differences in the PA
profiles at 3.1 and 1.4~GHz, which suggest new interpretations are
necessary. Recent theoretical advances in understanding the effects of
strong refraction in the pulsar magnetosphere \citep{pet00,wsve03}
have shown the potential of such effects on the total power profiles,
but the exact effects on polarisation are yet to be explored.

Also, there exists an interesting subset of pulsars with a virtually
flat PA profile. These pulsars show very high linear polarisation over
a wide frequency range and relatively flat spectral index. Many (but
not all) are high spin-down, young objects. Most have simple single
Gaussian profiles (e.g. PSR J1048$-$5832), occasionally two widely
separated components are seen (e.g. PSR J1302$-$6350).  Also, the
circular polarisation in these objects tends to have a single
handedness.  We concur with \citet{man96} that the components in these
young pulsars likely originate far from the magnetic pole.

\section*{Acknowledgments}
We would like to thank Alex Judge and the staff at the Parkes
telescope for help with the observations. The Australia Telescope is
funded by the Commonwealth of Australia for operation as a National
Facility managed by the CSIRO.
\label{lastpage}
\bibliographystyle{mn2e}
\bibliography{journals,modrefs,psrrefs,crossrefs}

\begin{thebibliography}{}

\bibitem[\protect\citeauthoryear{{Backer}, {Rankin} \& {Campbell}}{{Backer}
  et~al.}{1976}]{brc76}
{Backer} D.~C.,  {Rankin} J.~M.,    {Campbell} D.~B.,  1976, Nature, 263, 202

\bibitem[\protect\citeauthoryear{{Clemens} \& {Rosen}}{{Clemens} \&
  {Rosen}}{2004}]{cr04}
{Clemens} J.~C.,  {Rosen} R.,  2004, ApJ, 609, 340

\bibitem[\protect\citeauthoryear{Cordes \& Hankins}{Cordes \&
  Hankins}{1977}]{ch77}
Cordes J.~M.,  Hankins T.~H.,  1977, ApJ, 218, 484

\bibitem[\protect\citeauthoryear{Cordes, Rankin \& Backer}{Cordes
  et~al.}{1978}]{crb78}
Cordes J.~M.,  Rankin J.~M.,    Backer D.~C.,  1978, ApJ, 223, 961

\bibitem[\protect\citeauthoryear{{Crawford}, {Manchester} \&
  {Kaspi}}{{Crawford} et~al.}{2001}]{cmk01}
{Crawford} F.,  {Manchester} R.~N.,    {Kaspi} V.~M.,  2001, A. J., 122,
  2001

\bibitem[\protect\citeauthoryear{{Edwards}, {Bailes}, {van Straten} \&
  {Britton}}{{Edwards} et~al.}{2001}]{ebvb01}
{Edwards} R.~T.,  {Bailes} M.,  {van Straten} W.,    {Britton} M.~C.,  2001,
  MNRAS, 326, 358

\bibitem[\protect\citeauthoryear{{Edwards} \& {Stappers}}{{Edwards} \&
  {Stappers}}{2002}]{es02}
{Edwards} R.~T.,  {Stappers} B.~W.,  2002, A\&A, 393, 733

\bibitem[\protect\citeauthoryear{{Edwards} \& {Stappers}}{{Edwards} \&
  {Stappers}}{2003}]{es03b}
{Edwards} R.~T.,  {Stappers} B.~W.,  2003, A\&A, 410, 961

\bibitem[\protect\citeauthoryear{{Everett} \& {Weisberg}}{{Everett} \&
  {Weisberg}}{2001}]{ew01}
{Everett} J.~E.,  {Weisberg} J.~M.,  2001, ApJ, 553, 341


\bibitem[\protect\citeauthoryear{{Granet}, {Zhang}, {Forsyth},
{Graves}, {Doherty}, {Greene}, {James}, {Sykes}, {Bird}, {Sinclair},
{Moorey} \& {Manchester}}{{Granet} et~al.}{2004}]{gzfg+04} Granet C.,
Zhang H. Z., Forsyth A. R., Graves G. R., Doherty P., Greene K. J.,
James G. L., Sykes P., Bird T. S., Sinclair M. W., Moorey G., and
Manchester R. N., IEEE Antennas and Propagation Magazine, In Press.


\bibitem[\protect\citeauthoryear{Hamilton, McCulloch, Ables \&
  Komesaroff}{Hamilton et~al.}{1977}]{hmak77}
Hamilton P.~A.,  McCulloch P.~M.,  Ables J.~G.,    Komesaroff M.~M.,  1977,
  MNRAS, 180, 1

\bibitem[\protect\citeauthoryear{Han, Manchester, Xu \& Qiao}{Han
  et~al.}{1998}]{hmxq98}
Han J.~L.,  Manchester R.~N.,  Xu R.~X.,    Qiao G.~J.,  1998, MNRAS, 300, 373

\bibitem[\protect\citeauthoryear{{Hobbs}, {Faulkner}, {Stairs}, {Camilo},
  {Manchester}, {Lyne}, {Kramer}, {D'Amico}, {Kaspi}, {Possenti}, {McLaughlin},
  {Lorimer}, {Burgay}, {Joshi} \& {Crawford}}{{Hobbs} et~al.}{2004}]{hfs+04}
{Hobbs} G.,  {Faulkner} A.,  {Stairs} I.~H.,  {Camilo} F.,  {Manchester} R.~N.,
   {Lyne} A.~G.,  {Kramer} M.,  {D'Amico} N.,  {Kaspi} V.~M.,  {Possenti} A.,
  {McLaughlin} M.~A.,  {Lorimer} D.~R.,  {Burgay} M.,  {Joshi} B.~C.,
  {Crawford} F.,  2004, MNRAS, 352, 1439

\bibitem[\protect\citeauthoryear{{Hotan}, {van Straten} \&
  {Manchester}}{{Hotan} et~al.}{2004}]{hvm04}
{Hotan} A.,  {van Straten} W.,    {Manchester} R.~N.,  2004, PASA, 21, 302

\bibitem[\protect\citeauthoryear{Johnston, Lyne, Manchester, Kniffen, D'Amico,
  Lim \& Ashworth}{Johnston et~al.}{1992}]{jlm+92}
Johnston S.,  Lyne A.~G.,  Manchester R.~N.,  Kniffen D.~A.,  D'Amico N.,  Lim
  J.,    Ashworth M.,  1992, MNRAS, 255, 401

\bibitem[\protect\citeauthoryear{Johnston \& Romani}{Johnston \&
  Romani}{2003}]{jr03}
Johnston S.,  Romani R.~W.,  2003, ApJ, 590, L95

\bibitem[\protect\citeauthoryear{{Karastergiou} \& {Johnston}}{{Karastergiou}
  \& {Johnston}}{2004}]{kj04}
{Karastergiou} A.,  {Johnston} S.,  2004, MNRAS, 352, 689

\bibitem[\protect\citeauthoryear{{Karastergiou}, {Johnston}, {Mitra}, {van
  Leeuwen} \& {Edwards}}{{Karastergiou} et~al.}{2003}]{kjm+03}
{Karastergiou} A.,  {Johnston} S.,  {Mitra} D.,  {van Leeuwen} A.~G.~J.,
  {Edwards} R.~T.,  2003, MNRAS, 344, L69

\bibitem[\protect\citeauthoryear{{Karastergiou}, {Kramer}, {Johnston}, {Lyne},
  {Bhat} \& {Gupta}}{{Karastergiou} et~al.}{2002}]{kkj+02}
{Karastergiou} A.,  {Kramer} M.,  {Johnston} S.,  {Lyne} A.~G.,  {Bhat}
  N.~D.~R.,    {Gupta} Y.,  2002, A\&A, 391, 247

\bibitem[\protect\citeauthoryear{Kramer}{Kramer}{1994}]{kra94}
Kramer M.,  1994, A\&AS, 107, 527

\bibitem[\protect\citeauthoryear{Kramer, Xilouris, Jessner \& Wielebinski
  R.;~Timofeev}{Kramer et~al.}{1996}]{kxj+96}
Kramer M.,  Xilouris K.~M.,  Jessner A.,    Wielebinski R.;~Timofeev M.,  1996,
  A\&A, 306, 867

\bibitem[\protect\citeauthoryear{Lyne \& Manchester}{Lyne \&
  Manchester}{1988}]{lm88} Lyne A.~G., Manchester R.~N., 1988, MNRAS,
  234, 477

\bibitem[\protect\citeauthoryear{Lorimer, Yates, Lyne \&
  Gould}{Lorimer et~al.}{1995}]{lylg95} Lorimer D.~R., Yates J.~A.,
  Lyne A.~G., Gould D.~M., 1995, MNRAS, 273, 411

\bibitem[\protect\citeauthoryear{McCulloch, Hamilton, Manchester \&
  Ables}{McCulloch et~al.}{1978}]{mhma78}
McCulloch P.~M.,  Hamilton P.~A.,  Manchester R.~N.,    Ables J.~G.,  1978,
  MNRAS, 183, 645

\bibitem[\protect\citeauthoryear{McKinnon}{McKinnon}{1997}]{mck97}
McKinnon M.,  1997, ApJ, 475, 763

\bibitem[\protect\citeauthoryear{McKinnon \& Stinebring}{McKinnon \&
  Stinebring}{1998}]{ms98}
McKinnon M.,  Stinebring D.,  1998, ApJ, 502, 883

\bibitem[\protect\citeauthoryear{Manchester}{Manchester}{1996}]{man96}
 Manchester R.~N., 1996, in Johnston S., Walker M.~A., Bailes M., eds,
 {IAU} Colloquium 160, Pulsars: Problems and Progress, Astronomical
 Society of the Pacific, San Francisco, p. 193

\bibitem[\protect\citeauthoryear{Manchester, Hamilton \& McCulloch}{Manchester
  et~al.}{1980}]{mhm80}
Manchester R.~N.,  Hamilton P.~A.,    McCulloch P.~M.,  1980, MNRAS, 192, 153

\bibitem[\protect\citeauthoryear{Manchester \& Johnston}{Manchester \&
  Johnston}{1995}]{mj95}
Manchester R.~N.,  Johnston S.,  1995, ApJL, 441, L65

\bibitem[\protect\citeauthoryear{Manchester, Taylor \& Huguenin}{Manchester
  et~al.}{1975}]{mth75}
Manchester R.~N.,  Taylor J.~H.,    Huguenin G.~R.,  1975, ApJ, 196, 83

\bibitem[\protect\citeauthoryear{{Maron}, {Kijak}, {Kramer} \&
  {Wielebinski}}{{Maron} et~al.}{2000}]{mkkw00a} {Maron} O., {Kijak}
  J., {Kramer} M., {Wielebinski} R., 2000, A\&AS, 147, 195

\bibitem[\protect\citeauthoryear{{Melrose} \& {Luo}}{{Melrose} \&
  {Luo}}{2004}]{ml04a}
{Melrose} D.~B.,  {Luo} Q.,  2004, MNRAS, 352, 915

\bibitem[\protect\citeauthoryear{{Mitra} \& {Li}}{{Mitra} \& {Li}}{2004}]{ml04}
{Mitra} D.,  {Li} X.~H.,  2004, A\&A, 421, 215

\bibitem[\protect\citeauthoryear{Morris, Graham, Seiber, Bartel \&
  Thomasson}{Morris et~al.}{1981}]{mgs+81}
Morris D.,  Graham D.~A.,  Seiber W.,  Bartel N.,    Thomasson P.,  1981,
  A\&AS, 46, 421

\bibitem[\protect\citeauthoryear{{Petrova}}{{Petrova}}{2000}]{pet00}
{Petrova} S.~A.,  2000, A\&A, 360, 592

\bibitem[\protect\citeauthoryear{{Petrova}}{{Petrova}}{2001}]{pet01}
{Petrova} S.~A.,  2001, A\&A, 378, 883

\bibitem[\protect\citeauthoryear{Qiao, Manchester, Lyne \& Gould}{Qiao
  et~al.}{1995}]{qmlg95}
Qiao G.~J.,  Manchester R.~N.,  Lyne A.~G.,    Gould D.~M.,  1995, MNRAS, 274,
  572

\bibitem[\protect\citeauthoryear{Radhakrishnan \& Cooke}{Radhakrishnan \&
  Cooke}{1969}]{rc69a}
Radhakrishnan V.,  Cooke D.~J.,  1969, ApL, 3, 225

\bibitem[\protect\citeauthoryear{Rankin}{Rankin}{1983a}]{ran83}
Rankin J.~M.,  1983a, ApJ, 274, 333

\bibitem[\protect\citeauthoryear{Rankin}{Rankin}{1983b}]{ran83a}
Rankin J.~M.,  1983b, ApJ, 274, 359

\bibitem[\protect\citeauthoryear{Rankin}{Rankin}{1986}]{ran86}
Rankin J.~M.,  1986, ApJ, 301, 901

\bibitem[\protect\citeauthoryear{Stinebring, Cordes, Rankin, Weisberg \&
  Boriakoff}{Stinebring et~al.}{1984}]{scr+84}
Stinebring D.~R.,  Cordes J.~M.,  Rankin J.~M.,  Weisberg J.~M.,    Boriakoff
  V.,  1984, ApJS, 55, 247

\bibitem[\protect\citeauthoryear{Taylor, Manchester \& Lyne}{Taylor
  et~al.}{1993}]{tml93}
Taylor J.~H.,  Manchester R.~N.,    Lyne A.~G.,  1993, ApJS, 88, 529

\bibitem[\protect\citeauthoryear{van Ommen, D'Alesssandro, Hamilton \&
  McCulloch}{van Ommen et~al.}{1997}]{vdhm97}
van Ommen T.~D.,  D'Alesssandro F.~D.,  Hamilton P.~A.,    McCulloch P.~M.,
  1997, MNRAS, 287, 307

\bibitem[\protect\citeauthoryear{von Hoensbroech, Kijak \& Krawczyk}{von
  Hoensbroech et~al.}{1998}]{hkk98}
von Hoensbroech A.,  Kijak J.,    Krawczyk A.,  1998, A\&A, 334, 571

\bibitem[\protect\citeauthoryear{von Hoensbroech \& Xilouris}{von Hoensbroech
  \& Xilouris}{1997}]{hx97}
von Hoensbroech A.,  Xilouris K.~M.,  1997, A\&AS, 126, 121

\bibitem[\protect\citeauthoryear{{Wang}, {Johnston} \& {Manchester}}{{Wang}
  et~al.}{2004}]{wjm04}
{Wang} N.,  {Johnston} S.,    {Manchester} R.~N.,  2004, MNRAS, 351, 599

\bibitem[\protect\citeauthoryear{{Weltevrede}, {Stappers}, {van den Horn} \&
  {Edwards}}{{Weltevrede} et~al.}{2003}]{wsve03}
{Weltevrede} P.,  {Stappers} B.~W.,  {van den Horn} L.~J.,    {Edwards} R.~T.,
  2003, A\&A, 412, 473

\bibitem[\protect\citeauthoryear{Xilouris, Kramer, Jessner \&
  Wielebinski}{Xilouris et~al.}{1994}]{xkjw94}
Xilouris K.~M.,  Kramer M.,  Jessner A.,    Wielebinski R.,  1994, A\&A, 288,
  L17

\bibitem[\protect\citeauthoryear{Xilouris, Kramer, Jessner, Wielebinski \&
  Timofeev}{Xilouris et~al.}{1996}]{xkj+96}
Xilouris K.~M.,  Kramer M.,  Jessner A.,  Wielebinski R.,    Timofeev M.,
  1996, A\&A, 309, 481

\bibitem[\protect\citeauthoryear{Xilouris, Seiradakis, Gil, Sieber \&
  Wielebinski}{Xilouris et~al.}{1995}]{xsg+95} Xilouris K.~M.,
  Seiradakis J.~H., Gil J.~A., Sieber W., Wielebinski R., 1995, A\&A,
  293, 153 

\end{thebibliography}
\end{document}